\documentclass{aastex63}

\begin{document}

\title{Mitigating Bias in CMB B-modes from Foreground Cleaning Using a Moment Expansion}

\author{Danielle Sponseller}
\affiliation{Department of Physics \& Astronomy, Johns Hopkins University, Baltimore, MD 21218, USA}

\author{Alan Kogut}
\affiliation{Code 665, Goddard Space Flight Center, Greenbelt, MD 20771, USA}

\begin{abstract}
    One of the primary challenges facing upcoming CMB polarization experiments aiming to measure the inflationary B-mode signal is the removal of polarized foregrounds.
    The thermal dust foreground is often modeled as a single modified blackbody, however overly simplistic foreground models can bias measurements of the tensor-to-scalar ratio $r$.
    As CMB polarization experiments become increasingly sensitive, thermal dust emission models must account for greater complexity in the dust foreground while making minimal assumptions about the underlying distribution of dust properties within a beam.
    We use \textit{Planck} dust temperature data to estimate the typical variation in dust properties along the line of sight and examine the impact of these variations on the bias in $r$ if a single modified blackbody model is assumed.
    We then assess the ability of the moment method to capture the effects of spatial averaging and to reduce bias in the tensor-to-scalar ratio for different possible toy models of dust emission.
    We find that the expected bias due to temperature variations along the line of sight is significant compared to the target sensitivities of future CMB experiments, and that the use of the moment method could reduce bias as well as shed light into the distribution of dust physical parameters.
\end{abstract}

\keywords{
\href{http://astrothesaurus.org/uat/322}{Cosmic microwave background radiation (322)};
\href{http://astrothesaurus.org/uat/836}{Interstellar dust (836)}
}

\section{Introduction}
\label{sec:intro}

Many current and upcoming cosmic microwave background (CMB) experiments aim to measure the B-mode polarization of the CMB.
B-modes are predicted as a result of primordial gravitational waves generated during inflation, a theorized period of exponential expansion in the early universe.
Their detection would not only provide compelling evidence for inflation, but would also shed light into physics at energy scales orders of magnitude beyond those accessible to particle accelerators and would potentially provide the first evidence for quantum gravity \citep{Krauss2014}.

The primordial B-mode signal is extremely faint, with current upper limits on the tensor-to-scalar ratio of $r < 0.036$ \citep{Ade2021}.
Further complicating its detection, a number of astrophysical foregrounds impart a B-mode polarization signature which must be carefully modeled and subtracted to leave only the primordial B-mode signal.
In particular, the dominant foregrounds of synchrotron emission and thermal dust emission overwhelm the predicted B-mode signal across all frequencies, with synchrotron dominating at low frequencies 
($\lesssim 70$ GHz) and dust at high frequencies ($\gtrsim 70$ GHz).

These foregrounds are polarized as a result of their interaction with the galactic magnetic field.
Synchrotron emission is produced by the motion of charged particles through the
galactic magnetic field, 
and is polarized perpendicular to the magnetic field.
It is generally modeled as a power law in frequency, although more complex models can include a spectral curvature parameter or spectral breaking.
Dust grains absorb UV photons from the interstellar radiation field and re-emit thermal radiation in the far-infrared.
The grains tend to align themselves so that their short axis is parallel to the direction of the magnetic field, while they emit most efficiently along their long axis.
This results in emission with polarization perpendicular to the local magnetic field.
Variations in the orientation of the magnetic field along the line of sight can thus result in frequency decorrelation.

At millimeter or longer wavelengths, thermal dust emission is often modeled as a modified blackbody with a power law emissivity.
While this model gives a good approximation of dust emission at the frequencies commonly used for CMB observations, the true emission spectrum is expected to be more complicated.
The true dust emission will be affected by variations in both the dust grain properties and the intensity of the local interstellar radiation field.
With the unprecedented sensitivities of upcoming CMB polarization experiments, foregrounds must be modeled with greater accuracy to be able to accurately recover the faint primordial B-mode signal.

Several studies have shown that inaccuracies in modelling foregrounds can lead to a bias in the tensor-to-scalar ratio $r$ \citep{ArmitageCaplan2012, Fantaye2011, Kogut2016, Stompor2016, Hensley2018}.
\cite{ArmitageCaplan2012} consider a series of simulated input foreground spectra and various models, finding that a mismatch in the complexity of the true foregrounds and the models used can bias $r$ by as much as $\delta r \approx 0.03$.
An experiment such as PICO \citep{Hanany2019} has a sensitivity of $r \approx 1 \cdot 10^{-4}$, illustrating the need to allow sufficient complexity while minimizing the number of free parameters.

\cite{Remazeilles2016} consider the effect of using overly simplistic foreground models in the context of the proposed CMB satellite experiments \textit{LiteBIRD} \citep{Matsumura2014}, \textit{COrE} \citep{COrE-Collaboration2011}, \textit{COrE+}, \textit{PRISM} \citep{Andre2014}, \textit{EPIC} \citep{Bock2008}, and \textit{PIXIE} \citep{Kogut2011}.
Two possible dust models were considered for the input sky maps -- a single modified blackbody model, and a two-component modified blackbody model.
A mismatch between the foreground model and the simulations was introduced by assuming a single component for the dust model.
Even when excluding frequencies above 360 GHz to minimize the impact of inaccurate modeling, the tensor-to-scalar ratio was still strongly biased by $3\sigma$ or more for all considered experimental configurations.
In most cases, poor $\chi^2$ values signaled the unsuccessful separation of CMB B-modes and foregrounds, however experiments with lower sensitivity and fewer high-frequency channels may not be able to detect a poor fit from the $\chi^2$ value.

The focus of this paper is on thermal dust emission.
Our goal is to place bounds on the possible dust complexity and assess its expected impact on bias in the tensor-to-scalar ratio.
In section \ref{sec:expected_bias} we analyze \textit{Planck} thermal dust maps to estimate the variation in dust temperature and spectral index along the line of sight.
We then estimate the bias in the tensor-to-scalar ratio as a function of standard deviation assuming a Gaussian distribution in temperature or spectral index and using a single modified blackbody model.
In section \ref{sec:moment_method} we perform component separation simulations within a single pixel utilizing a dust parameterization known as the moment method \citep{Chluba2017, Rotti2020, Remazeilles2020, Vacher2022}, and assess its flexibility in modeling the effects of a continuous distribution of dust parameters and reducing bias.
Finally, we discuss what we could learn about dust properties from the moments returned by this method in section \ref{sec:maxent}.
We present our conclusions in \ref{sec:conclusions}.

\begin{figure}
    \centering
    \includegraphics[width = 0.7\columnwidth]{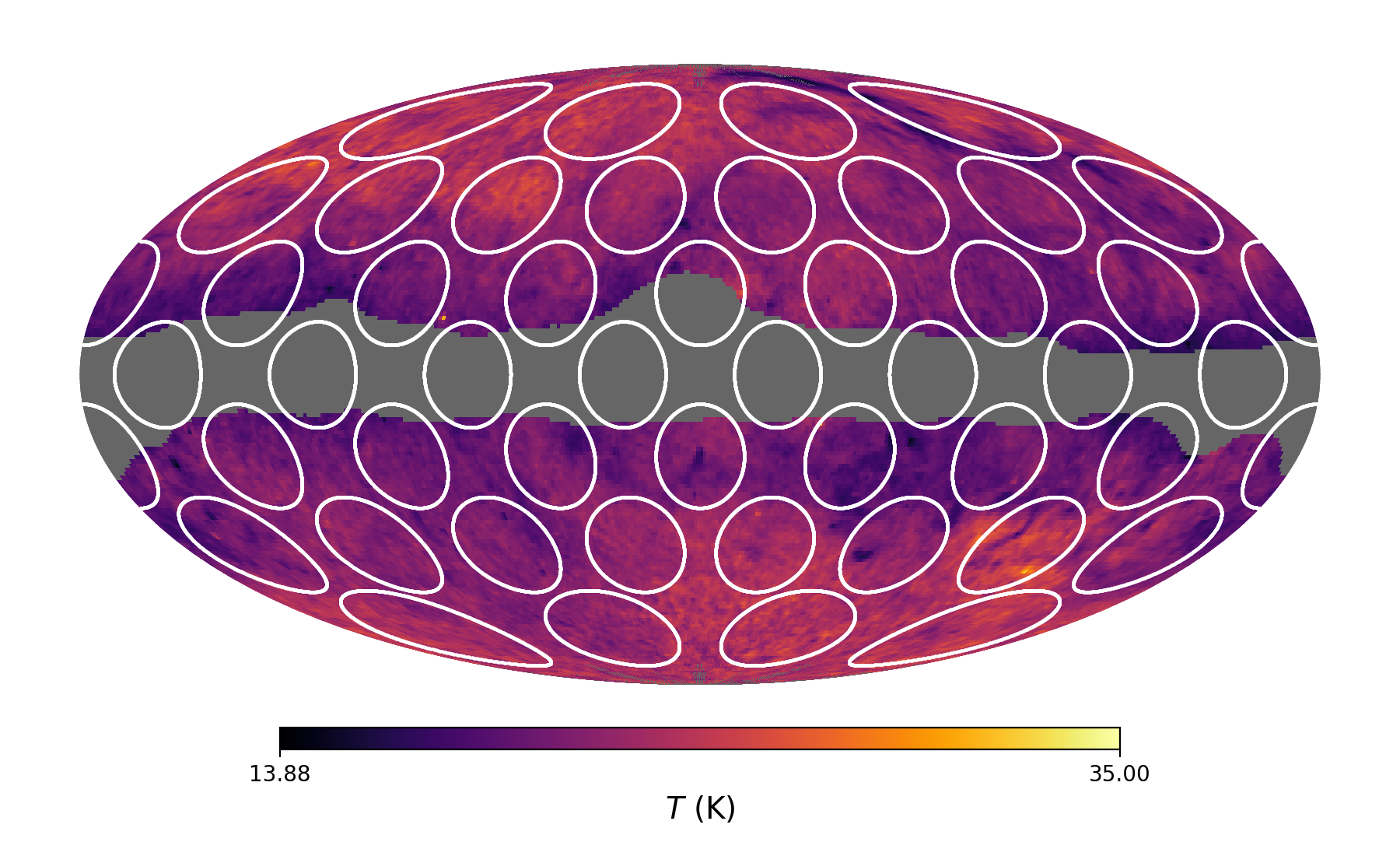}
    \caption{A \textit{Planck} dust temperature map overlaid with disks of equal radius based on a HEALPix grid with $N_{side} = 2$. The disk radius shown here is 12.5 degrees. The galactic plane is masked out (indicated by the gray region).}
	\label{fig:map_disks}
\end{figure}

\section{Expected bias due to dust variations}
\label{sec:expected_bias}

Each beam contains a superposition of multiple modified blackbody spectra, leading to observed dust temperature and spectral index variations across the sky.
However, even in the limit of a pencil beam variations along the line of sight are likely to result in spatial averaging.
As CMB experiments become increasingly sensitive, thermal dust emission models must be improved to capture dust complexity along the line of sight to avoid biasing the tensor-to-scalar ratio.

\subsection{Estimated line-of-sight variations in dust properties}
\label{sec:expected_var}

Observations show that the dust physical parameters vary when looking in different directions on the sky, so it is likely that similar variations occur along the line of sight.
To place bounds on the dust complexity along the line of sight, we analyze \textit{Planck} thermal dust maps produced using the Commander algorithm \citep{Eriksen2004} and use variations in the transverse direction as a proxy.
We quantify these variations by dividing the sky into disks and measuring the standard deviations in temperature and spectral index as a function of disk radius.
This is illustrated in figure \ref{fig:map_disks}, where we show a \textit{Planck} dust temperature map overlaid with sample disks of radius 12.5 degrees.
The grey region shows a sample mask, removing 20\% of the sky near the galactic plane.

Disk centers and radii are chosen by using HEALPix grids of various resolution orders and setting the disk radius to the minimum distance from any cell center to its boundary to avoid any overlap.
The disks shown in figure \ref{fig:map_disks} correspond to a HEALPix grid with $N_{side}=2$.
We then compute the standard deviation within each disk using all pixels from the full-resolution map ($N_{side}=256$) contained within that disk.
We note that no apparent dependence is observed between the galactic latitude and the standard deviation in either parameter.
The results for each disk are then averaged to estimate the expected variation in temperature and spectral index along the line of sight.

\begin{figure}
    \centering
    \includegraphics[width = 0.49\columnwidth]{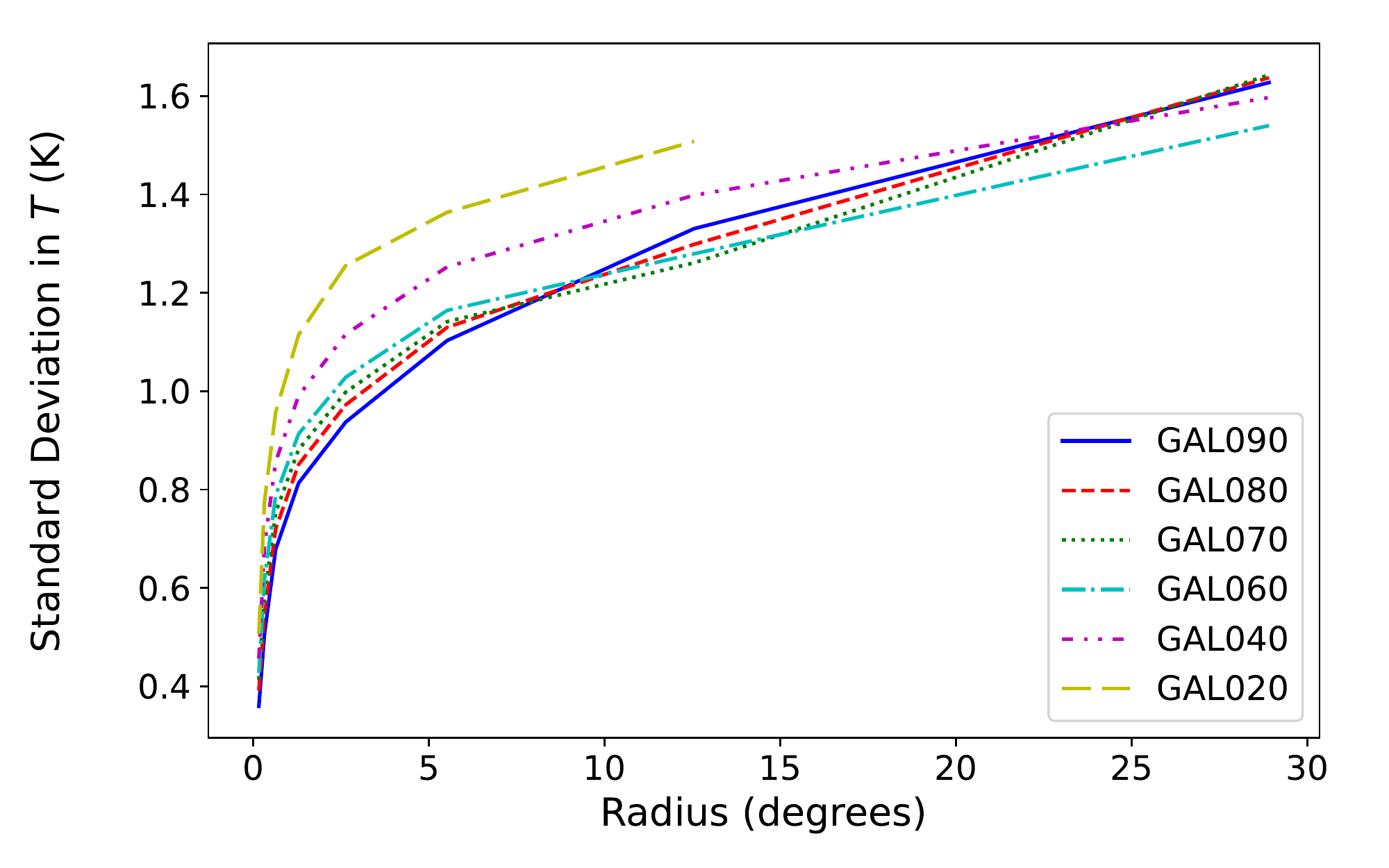}
    \includegraphics[width = 0.49\columnwidth]{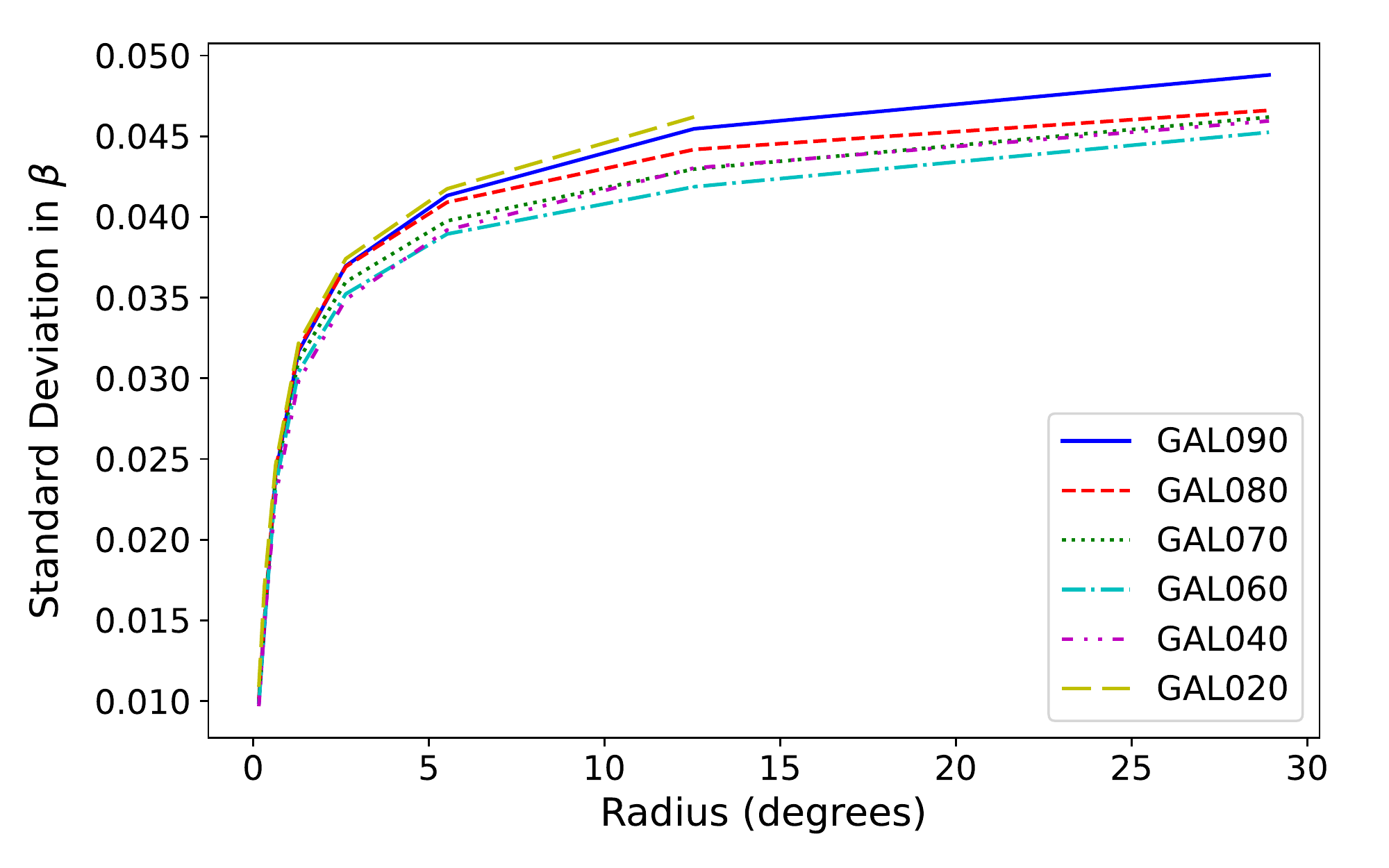}
    \caption{The mean standard deviation in temperature (left) and spectral index (right) as a function of disk radius for several sky masks. The GAL090 mask removes all but the cleanest 90\% of the sky, GAL080 the cleanest 80\%, etc. The choice of mask does not significantly affect the results.}
	\label{fig:std_radius}
\end{figure}

This was repeated using various \textit{Planck} sky masks.
A standard deviation was calculated for any disk for which more than half of its pixels were unmasked, and only those pixels were used in the calculation.
The standard deviations in temperature and spectral index as a function of disk radius are shown in figure \ref{fig:std_radius} for each considered sky mask.
The standard deviations appear to be nearly independent of the chosen masks.
In both cases, the standard deviation increases as a function of radius and begins to flatten out significantly beyond a radius of about ten degrees.

Without direct measurements of the line-of-sight dust complexity, we use variations in the transverse direction as a proxy for those along the line of sight.
We want to choose a disk radius such that dust complexity in the transverse direction is similar to that along the line of sight.
Near the galactic poles, the simplest geometry (transverse distance equal to line of sight distance) would suggest a fairly large disk radius.
Since figure \ref{fig:std_radius} shows that the standard deviation flattens out after about ten degrees, for simplicity we adopt a disk radius of 30 degrees throughout the remainder of this paper.
For a disk radius of 30 degrees and considering the cleanest 80\% of the sky, the expected standard deviations in temperature and spectral index are approximately 1.6 K and 0.045, respectively.

\subsection{Impact of dust variations on bias in the tensor-to-scalar ratio}

Several studies have demonstrated that a mismatch in the complexity of the true dust emission spectrum and the model used can result in a significant bias on the tensor-to-scalar ratio $r$
\citep{ArmitageCaplan2012, Fantaye2011, Kogut2016, Stompor2016, Hensley2018}.
The goal of this section is to directly quantify the relationship between the complexity of the true dust relative to the model and the resulting bias in $r$.
To do this, we generate dust and synchrotron spectra and simulate component separation using an overly simplistic dust model.

Synchrotron emission is often modeled as a power law, with a spectrum given by
\begin{equation}
    I_s(\nu) = \epsilon_{s,0} \left( \frac{\nu}{\nu_0} \right)^{\beta_s}
    \label{eqn:synch_mbb}
\end{equation}
where $\epsilon_{s,0}$ is the emissivity at reference frequency $\nu_0$ and $\beta_s = -1.1$ is the synchrotron spectral index \citep{Planck-Collaboration2020a}.
Note that here we are using units of surface brightness.
For this test, the input synchrotron spectrum exactly matches the complexity of the model used.

Thermal dust emission is generally modeled as a single modified blackbody, with a spectrum given by
\begin{equation}
    I_d(\nu) = \epsilon_{d,0} \left( \frac{\nu}{\nu_0} \right)^{\beta_d} B_\nu(T_d)
    \label{eqn:dust_mbb}
\end{equation}
where $\epsilon_{d,0}$ is the emissivity at reference frequency $\nu_0$, $\beta_d = 1.53$ is the dust spectral index, and $B_\nu(T_d)$ is the Planck blackbody spectrum corresponding to the dust temperature $T_d = 19.6$ K \citep{Planck-Collaboration2020a}.
Realistically however, within any given beam there will be a superposition of dust spectra at various temperatures and spectral indices.
For this test, the simulated dust consists of a superposition of modified blackbody spectra with either a Gaussian distribution in temperature and a constant spectral index:
\begin{equation}
    I(\nu) = \int e^{-\frac{1}{2} \left(\frac{T - T_d}{\sigma_T}\right)^2} \left( \frac{\nu}{\nu_{\textrm{ref}}} \right)^{\beta_d} B_\nu(T) \,dT
    \label{eqn:gauss_T}
\end{equation}
or a Gaussian distribution in spectral index with a constant temperature:
\begin{equation}
    I(\nu) = \int e^{-\frac{1}{2} \left(\frac{\beta - \beta_d}{\sigma_\beta}\right)^2} \left( \frac{\nu}{\nu_{\textrm{ref}}} \right)^{\beta} B_\nu(T_d) \,d\beta
    \label{eqn:gauss_beta}
\end{equation}
where $T_d = 19.6$K, $\beta_d = 1.53$, and $\sigma_T$ and $\sigma_\beta$ are the standard deviations of temperature and spectral index, respectively.
We model dust as only a single modified blackbody and vary the width of the Gaussian distribution to observe the relationship between the standard deviation and the resulting bias in $r$.
Input spectra are generated in the PICO frequency bands, using noise scaled by $10^{-6}$ to perform an approximately noiseless simulation while still preserving the relative frequency weighting of the PICO channels.
No CMB component is included so that the best-fit CMB amplitude for each fit directly represents the bias.
Note that we are not fitting multipoles, but rather fitting an SED within an individual line of sight. A Markov Chain Monte Carlo (MCMC) method is used to perform the fit.

\begin{figure}
    \centering
    \includegraphics[width = 0.5\columnwidth]{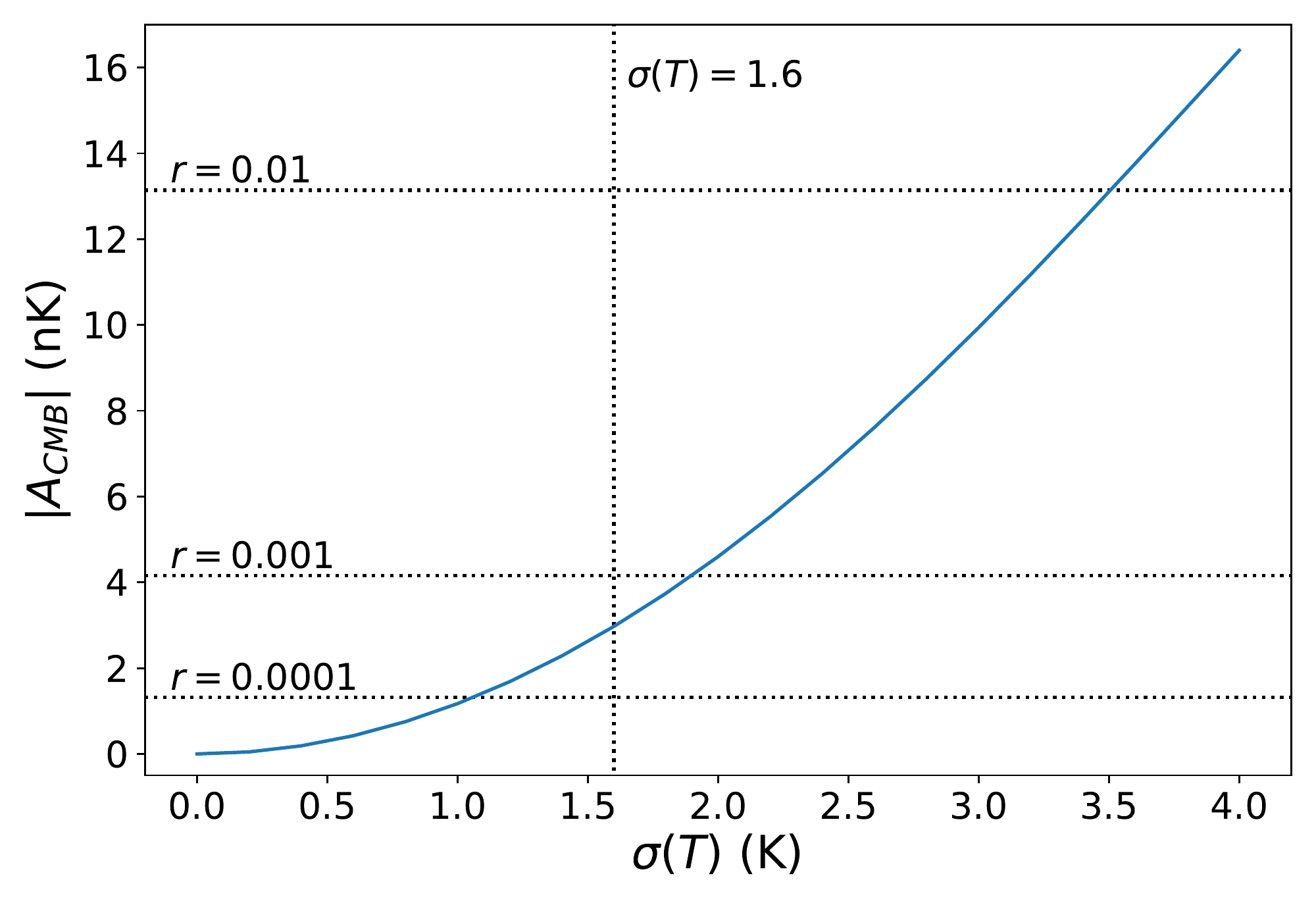}
    \caption{The bias in the CMB component as a function of standard deviation in dust temperature when a single modified blackbody model is used to fit the dust component in the PICO frequency bands. The corresponding $r$ values are shown for $l \in$ [80, 100] and a vertical line shows the expected standard deviation in temperature found in section \ref{sec:expected_var}.}
	\label{fig:bias_sigma_temp}
\end{figure}

Bias due to foreground complexity depends not only on the intrinsic dust variations, but also on the frequency channels used in a given experiment.
Here we evaluate spectra in the PICO frequency bands, with 21 frequency channels ranging from 21 to 800 GHz.
First, we consider dust with a Gaussian distribution of temperatures and uniform spectral index along the line of sight.
The synchrotron spectrum is described by a power law with a spectral index $\beta_s = -1.1$ and is normalized to the \textit{Planck} polarized intensity at 30 GHz.
The dust has a spectral index $\beta_d = 1.53$ and a central temperature $T_{d,0} = 19.6$ K, and the standard deviation of the dust temperature distribution $\sigma_T$ is varied between $0$ and $4$ K.
Each simulated spectrum was fit with a model containing dust, synchrotron, and CMB components, where the dust is modeled as a single modified blackbody, the synchrotron emission is modeled as a power law so that the complexity of the synchrotron model exactly matches the true complexity, and the CMB component is fit as an amplitude in nK.

\begin{figure}
    \centering
    \includegraphics[width = 0.5\columnwidth]{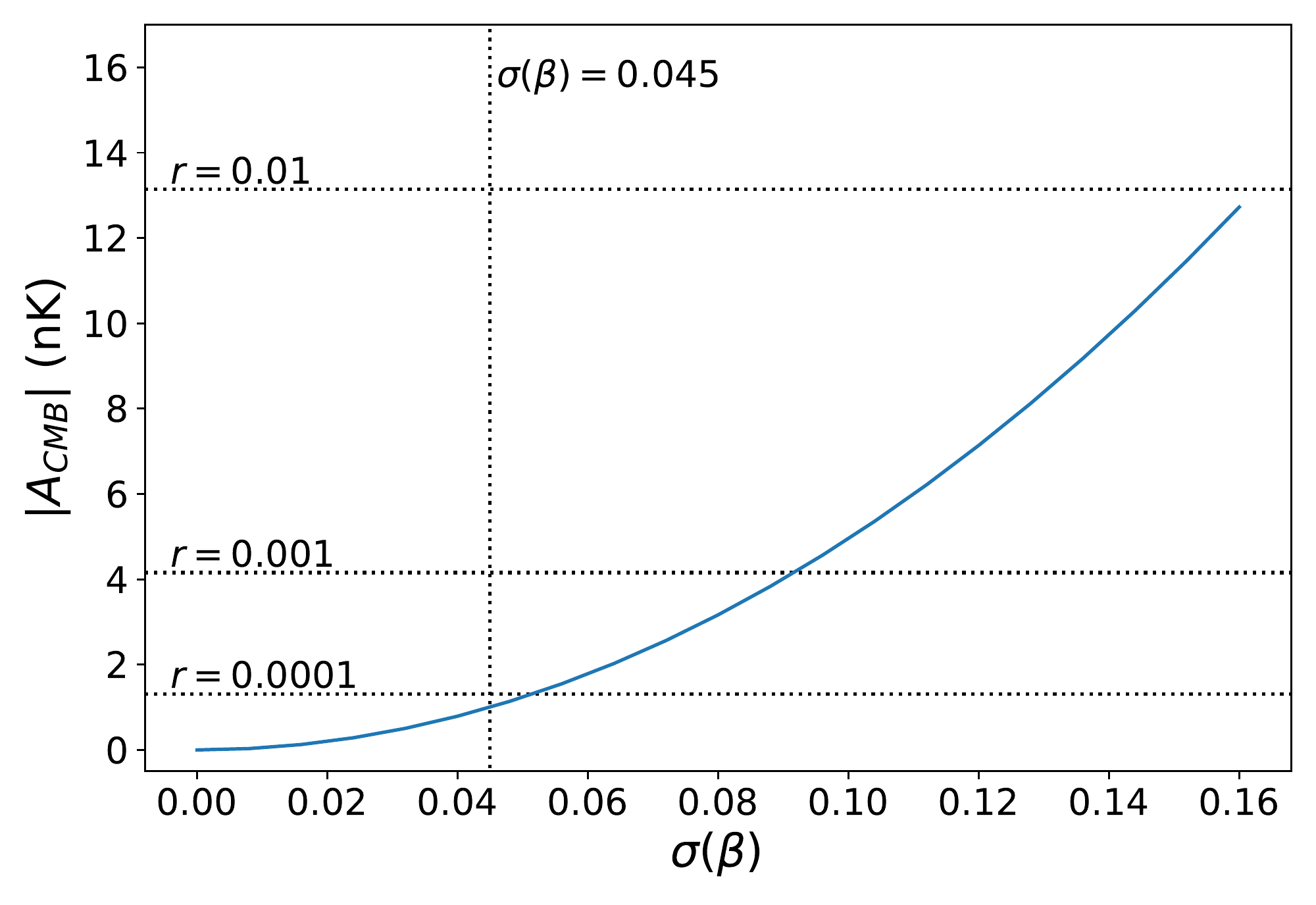}
    \caption{The bias in the CMB component as a function of standard deviation in dust spectral index when a single modified blackbody model is used to fit the dust component in the PICO frequency bands. The corresponding $r$ values are shown for $l \in$ [80, 100] and a vertical line shows the expected standard deviation in spectral index found in section \ref{sec:expected_var}.}
	\label{fig:bias_sigma_beta}
\end{figure}

Figures \ref{fig:bias_sigma_temp} and \ref{fig:bias_sigma_beta} show the bias in the CMB amplitude as a function of standard deviation in temperature and spectral index, respectively.
Since we are only interested in determining the bias in CMB B-modes, we should only consider the B-mode component of foreground emission.
We normalize our simulated foregrounds using \textit{Planck} total power, which includes both E- and B-modes.
These components do not have equal weight; approximately one third of the total power from polarized foregrounds comes from B-modes \citep{Planck-Collaboration2016b}.
We take this into account by scaling the bias by $1/\sqrt{3}$, finding that a standard deviation in temperature of 1.6 K corresponds to a bias in $r$ of approximately $1.4 \cdot 10^{-4}$ and a standard deviation in spectral index of 0.045 corresponds to a bias in $r$ of approximately $2 \cdot 10^{-5}$.
While the expected bias due to spectral index variations is negligible, that due to temperature variations will become significant for future CMB experiments such as PICO, illustrating the need for improved dust models.

\section{Fitting foregrounds using the moment method}
\label{sec:moment_method}

Several toy models of dust emission have been developed which account for possible distributions of the dust temperature and spectral index, however the distribution of dust properties is not well-understood.
Component separation methods must be capable of capturing complex variations in the dust emission spectrum with minimal prior knowledge of the true distribution of dust properties within a given pixel.

\subsection{Dust Emission Models}

Thermal dust emission is often modeled as a modified blackbody.
However, the size, composition, temperature, and spectral index of dust grains all affect the resulting emission spectrum and are expected to vary along the line of sight.
A number of empirical models have been developed which consider variations in these properties.

Dust consists largely of carbonaceous and silicate components, however it is not known whether these components exist primarily as separate species or composite particles.
\cite{Finkbeiner1999} consider a model with two components representing distinct dust species, motivated in part by \cite{Pollack1994}, finding that this model produces a significantly better fit to observational data than a single modified blackbody model.
However, if dust exists as distinct carbonaceous and silicate species, the temperatures of these two populations would be expected to differ.
This would result in a fractional polarization which varies with frequency, while observations show a relatively flat fractional polarization spectrum \citep{Ashton2018}.
\cite{Draine2021} present an alternative dust model, \textit{astrodust}, in which dust is represented by a single population consisting of larger composite grains formed through coagulation along with nanoparticles resulting from grain fragmentation.
Because it is a single population of dust particles, this picture of dust in the interstellar medium (ISM) is more consistent with observations of the polarization fraction.
\cite{Guillet2018} propose a dust model compatible with \textit{Planck} observations consisting of polycyclic aromatic hydrocarbons (PAHs), amorphous carbon, and astrosilicates.

While a two-component model can reproduce some of the observed flattening of the dust emission spectrum, it assumes two distinct dust species with fixed spectral indices.
\cite{Paradis2011} present a two-level systems (TLS) model which successfully reproduces observed variations in the dust spectral index without needing to divide the dust into separate populations.
This model considers the internal structure of amorphous silicate grains, resulting in a spectral index that depends on both temperature and frequency.

The size of dust grains also affects their thermal emission.
Very small grains have heat capacities that are small compared to the energy imparted by a single UV photon, resulting in a spike in temperature followed by a cool-down process in which photons are emitted through vibrational transitions \citep{Meny2007}.
The temperature distribution of these very small grains is described by the transient heating model, which has a constant spectral index and a temperature distribution with a peak at low temperatures and a tail to high temperatures.

\begin{figure}
    \centering
    \includegraphics[width = 0.6\columnwidth]{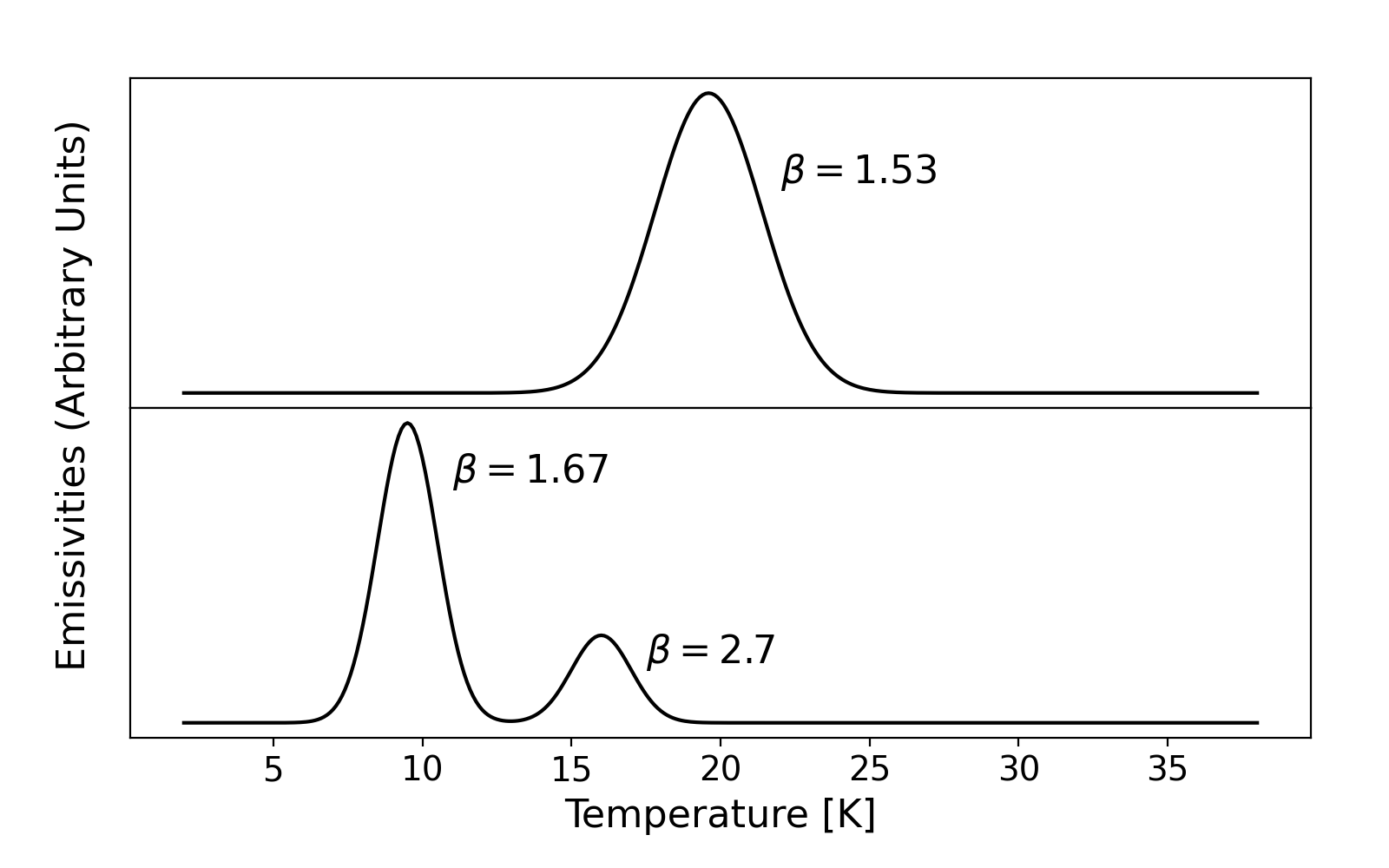}
    \caption{The temperature distributions and spectral indices used for the one-Gaussian (top) and two-Gaussian (bottom) models.}
	\label{fig:model_emissivities}
\end{figure}

\begin{figure}
    \centering
    \includegraphics[width = 0.6\columnwidth]{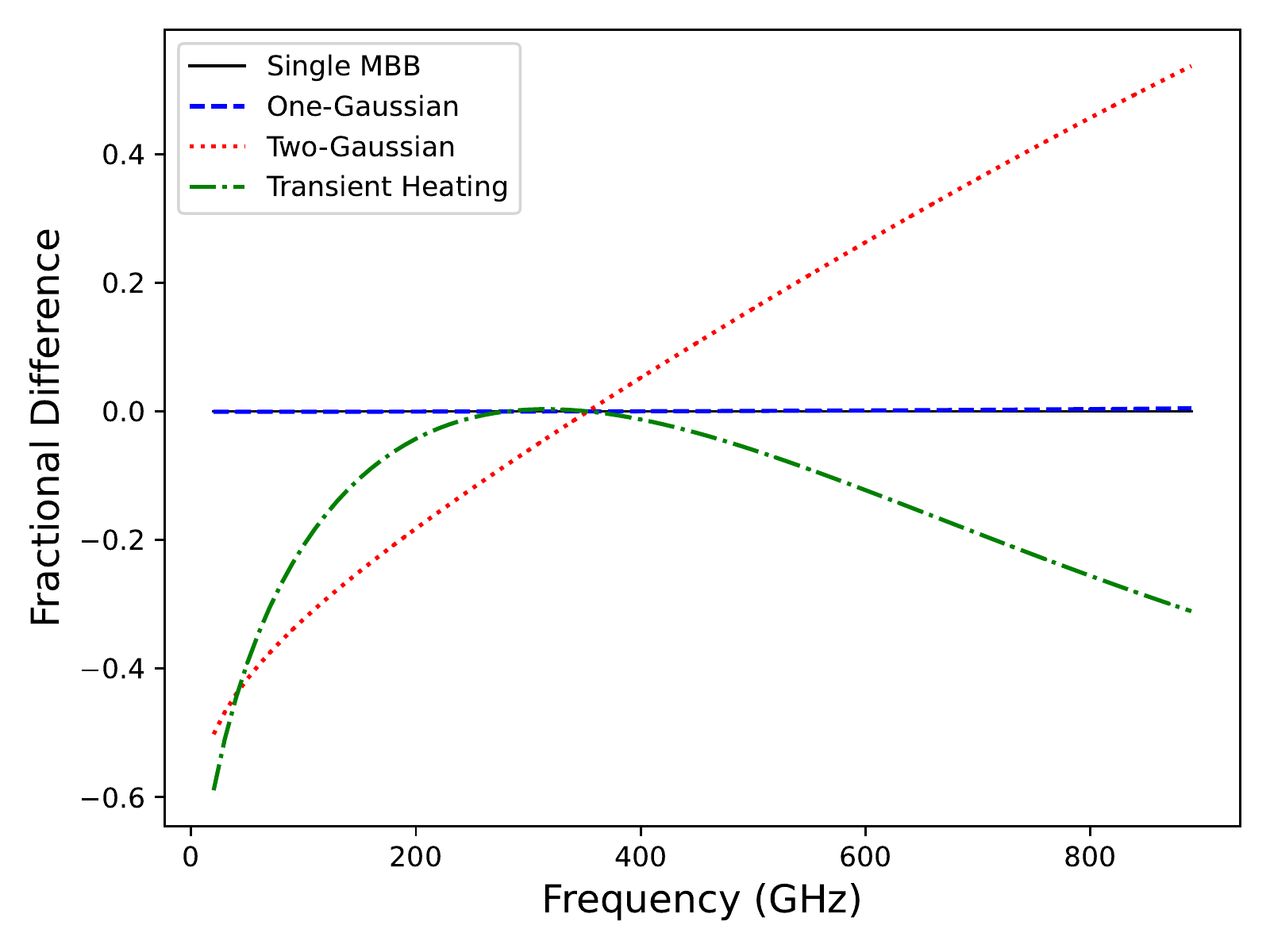}
    \caption{The fractional difference between each of the toy models considered and a single modified blackbody spectrum. The spectra are normalized at 353 GHz.}
	\label{fig:model_diffs}
\end{figure}

It is important to note that our goal is not to reproduce physical dust properties in detail, but rather to capture broad features due to dust complexity along the line of sight.
We start by considering a model with a Gaussian distribution of temperatures with either one or two peaks and a constant spectral index.
Figure \ref{fig:model_emissivities} shows the emissivity distributions for the one-Gaussian and two-Gaussian models.
The width of the one-Gaussian model is chosen to reflect the expected standard deviation in dust temperature found in section \ref{sec:expected_bias}.
The central temperature and spectral index for the one-Gaussian model are taken from a \textit{Planck} single-component fit, while the central temperatures and spectral indices for the Two-Gaussian model are taken from the \cite{Finkbeiner1999} two-component fit.
We then adopt the transient heating model as an example of a broad temperature distribution.
Figure \ref{fig:model_diffs} shows the fractional difference between each of these dust emission models after they have been normalized to the \textit{Planck} polarized intensity at 353 GHz.

It is important to choose a parameterization which allows sufficient complexity without making overly restrictive assumptions about the dust emission.
The moment method, introduced by \cite{Chluba2017}, expands the fundamental dust spectral energy distribution (SED) in terms of the dust physical parameters to obtain spectral basis functions characteristic of the effects of spatial averaging.

\subsection{Dust Parameterization}

Considering only variations along the line of sight, the observed emission spectrum is represented by a superposition of modified blackbody spectra with unique amplitude, temperature, and spectral index at each point along the line of sight:
\begin{equation}
    I_\nu = \int I_\nu(A(r), \beta(r), T(r)) dr.
\end{equation}
In the moment method, the amplitudes corresponding to each $(\beta, T)$ can be thought of as describing a probability density distribution in temperature and spectral index $\rho(\beta, T)$ \citep{Rotti2020}:
\begin{equation}
    I_\nu = \int \rho(\beta, T) I_\nu(A_0, \beta, T) d\beta dT
\end{equation}
where $A_0$ is a constant reference amplitude.
A multidimensional Taylor expansion of the fundamental spectral energy distribution (SED) is then performed around the average parameters $\bar{p}$, producing a set of spectral basis functions consisting of derivatives and cross-derivatives with respect to the dust physical parameters.
The model parameters to be optimized are then the central moments of the dust temperature and spectral index, providing direct insight into the nature of the dust.

\begin{figure}
    \centering
    \includegraphics[width = 0.49\columnwidth]{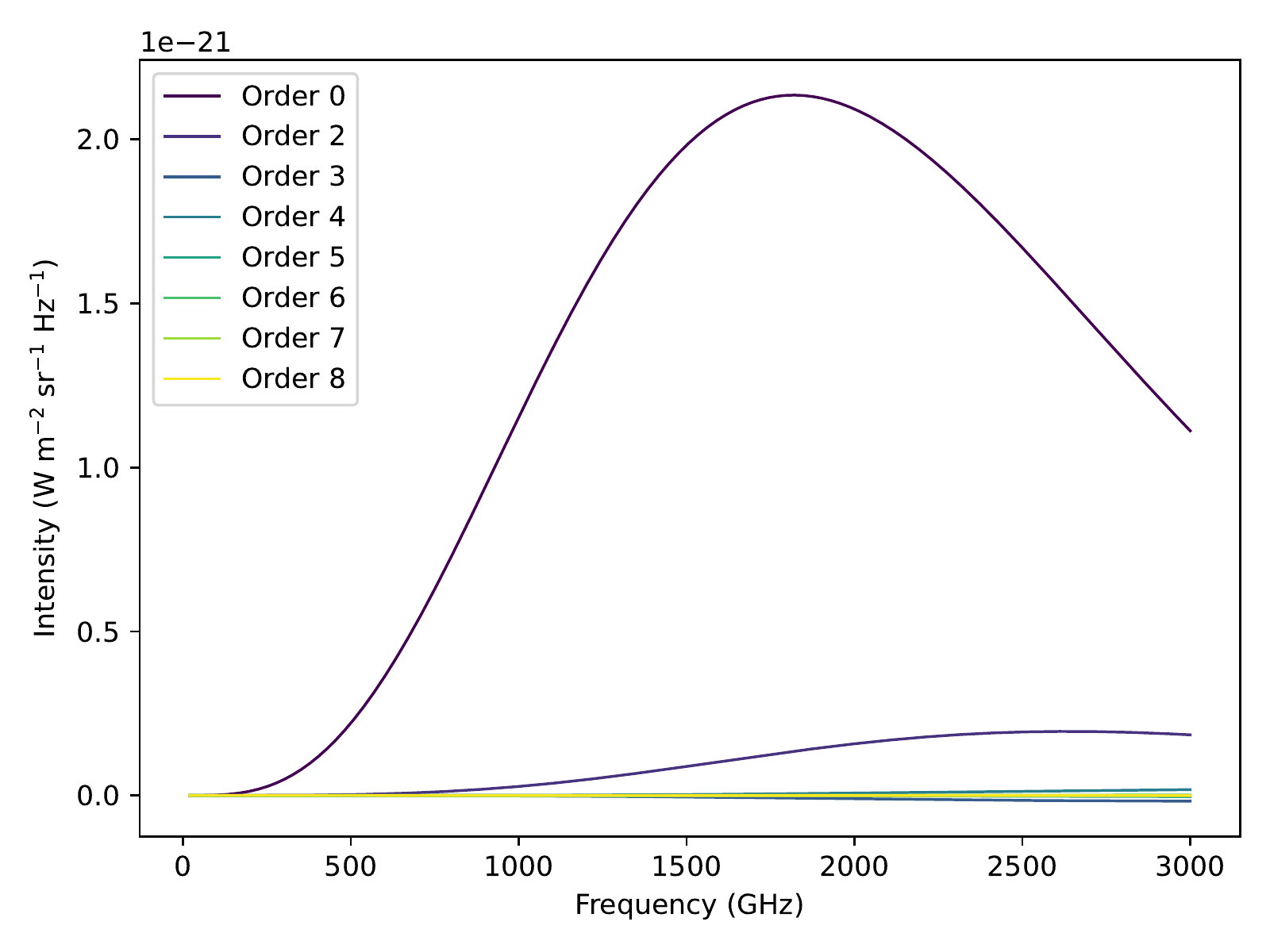}
    \includegraphics[width = 0.49\columnwidth]{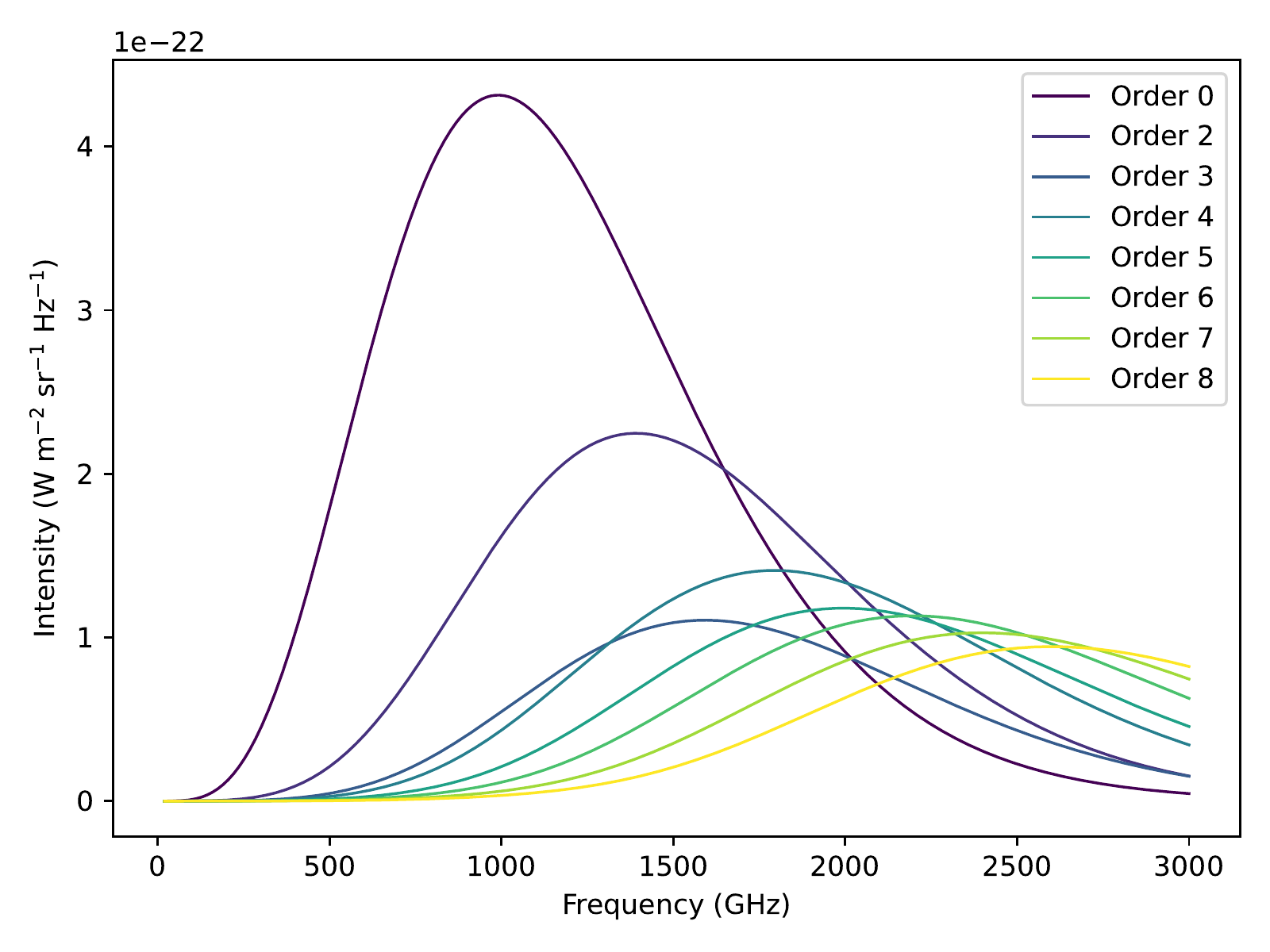}
    \caption{Contribution of each expansion order for two different dust models. The Gaussian model (left) is dominated by the first two expansion orders, while the transient heating model (right) requires a higher-order expansion.}
	\label{fig:mm_basis}
\end{figure}

The fundamental SED describing each emitting element along the line of sight is given by a modified blackbody spectrum:
\begin{equation}
    I_\nu(A, \beta, T) = A \left( \frac{\nu}{\nu_0} \right)^{\beta} B_\nu(T).
\end{equation}
All derivatives with respect to the dust physical parameters commute.
Following \cite{Chluba2017}, we expand with respect to the inverse of the dust temperature $\tau \equiv T^{-1}$ as computing the SED derivatives is more straightforward in this parameterization.
The spectral basis functions then have the form:
\begin{equation}
    I_{m,n}(\nu, \bar{p}) =
    \bar{A} \left( \frac{\nu}{\nu_0} \right)^{\bar{\beta}} \ln^m \left( \frac{\nu}{\nu_0} \right)
    \frac{\partial^n}{\partial^n \tau} B_\nu \left( \frac{1}{\bar{\tau}} \right)
\end{equation}
for $m$ derivatives with respect to $\beta$ and $n$ derivatives with respect to $\tau$.
Similarly, the central moments are given by:
\begin{equation}
    \omega_{m,n} = \left \langle (\beta - \bar{\beta})^m (\tau - \bar{\tau})^n \right \rangle
\end{equation}
Figure \ref{fig:mm_basis} shows the contribution of each order of the expansion to the spatially averaged spectrum.
As in \cite{Chluba2017}, the first-order term is kept at zero to fix the average parameters and thus is not shown.
For a Gaussian distribution of dust temperatures, the first two terms dominate the expansion, however dust with a tail to high temperatures requires several moments to be accurately represented.

\subsection{Implementation}

We now want to test the moment method on simulated spectra based on the toy models described above to determine whether fitting the dust component with this parameterization can reduce the expected bias in the tensor-to-scalar ratio and to assess what future CMB experiments could tell us about the distribution of dust properties.

One challenge in using the moment method is that the number of parameters can increase rapidly with increasing expansion order.
A recent analysis by \cite{Osumi2021} finds no significant variation in the dust spectral index, suggesting temperature variations may have a greater impact on spatial averaging of dust SEDs.
The results described in section 2.2 also suggest that the expected bias due to line-of-sight variations in spectral index is lower than that expected from temperature variations.
To reduce the number of free parameters, we therefore only expand the modified blackbody spectrum in terms of $\tau$.
In this case, incrementing the expansion order adds only one additional parameter.
The spectral basis functions are then:
$$ I_{0,n}(\nu, \bar{p}) =
\bar{A} \left( \frac{\nu}{\nu_0} \right)^{\bar{\beta}} \frac{\partial^n}{\partial^n (\tau)} B_\nu \left( \frac{1}{\bar{\tau}} \right) $$
and the model parameters are the statistical moments of $\tau$.

Simulated dust spectra are obtained by generating two-dimensional opacity distributions in temperature and spectral index according to the chosen model and calculating the spectrum produced by the superposition of each component in the PICO frequency bands.
The resulting spectrum is then normalized to the Planck polarized dust intensity at 353 GHz.
The true moments are calculated according to the equation below, where $A(\beta, \tau)$ represents the input dust opacity at each $(\beta, \tau)$.
$$ \omega_{0,n} =
\frac{1}{A_{tot}} \sum_{\beta, \tau} A(\beta, \tau) (\tau - \bar{\tau})^n $$
Derivatives with respect to $\tau$ are calculated numerically for each set of parameters.

\subsection{Fitting Methods}

The \textsc{emcee}\footnote{\url{https://emcee.readthedocs.io/en/stable/}} \citep{Foreman-Mackey2013} python package was used to fit simulated spectra using an ensemble Markov Chain Monte Carlo (MCMC) approach.
The dust component was modeled using the moment method.
To improve the performance, a zeroth-order fit was performed initially and the expansion order was incremented until the target order was reached, using the previous fits and uncertainties to initialize the walkers at each step.

The convergence of the distribution was monitored by tracking the integrated autocorrelation time for each parameter.
The chains were set to evolve until either the chains converged (i.e. the number of iterations reached 100 times the maximum autocorrelation time) or a set maximum number of iterations ($N_{\textrm{max}} = 10^{6}$) was reached.
In the results presented below, the MCMC chains fully converged in all cases.

\subsection{Results}
\label{sec:results}

To assess the performance of this method, we generate noiseless simulated spectra containing only dust emission using the PICO frequency channels.
We then perform multiple tests fitting both dust and CMB components at each expansion order to determine the order required to avoid significant bias in the tensor-to-scalar ratio.
We perform noiseless tests and use the root-mean-square fractional residual as the goodness-of-fit.
For each expansion order, 20 simulated fits are performed.

\begin{figure}
    \centering
    \includegraphics[width = 0.32\textwidth]{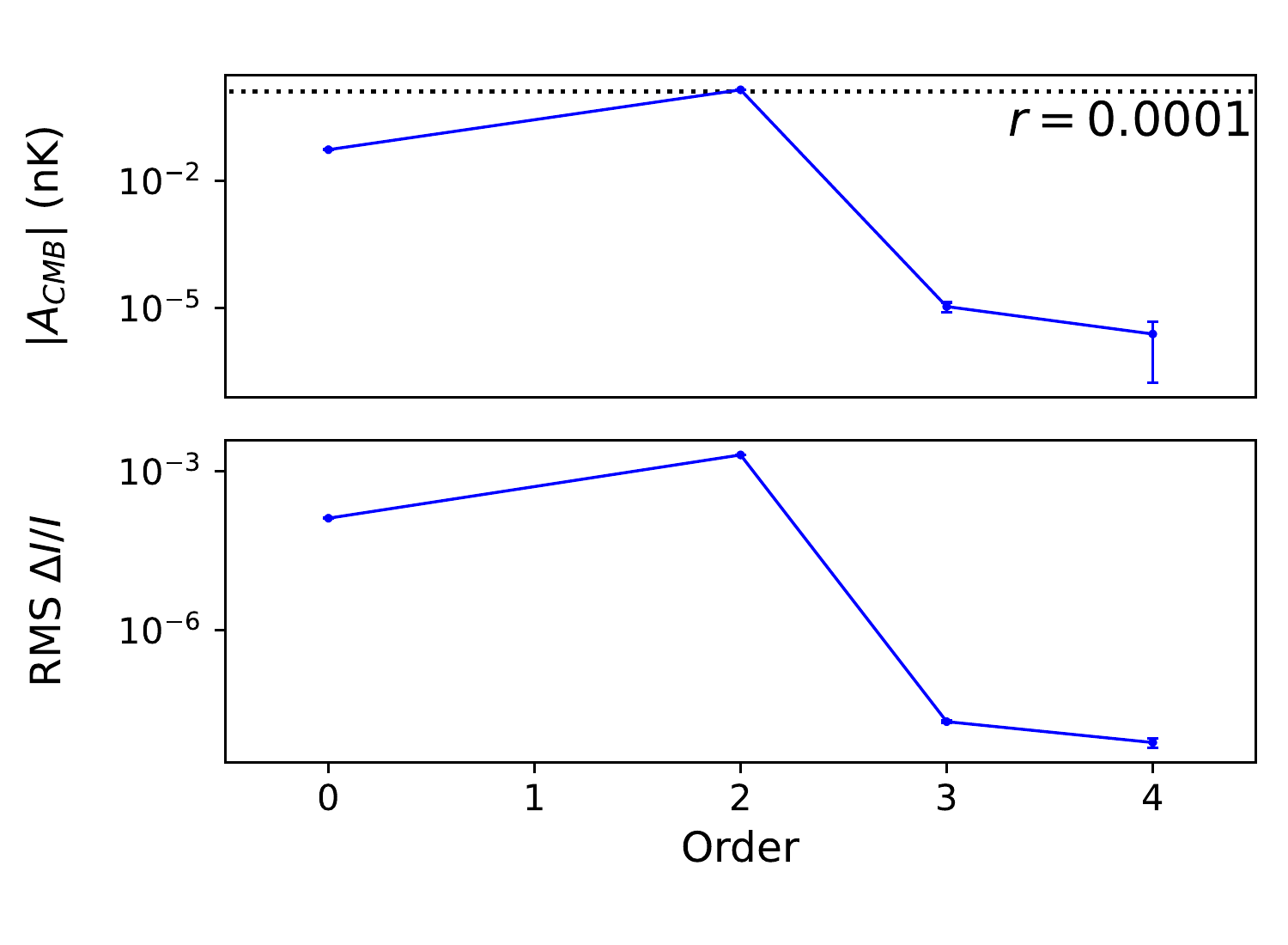}
    \includegraphics[width = 0.32\textwidth]{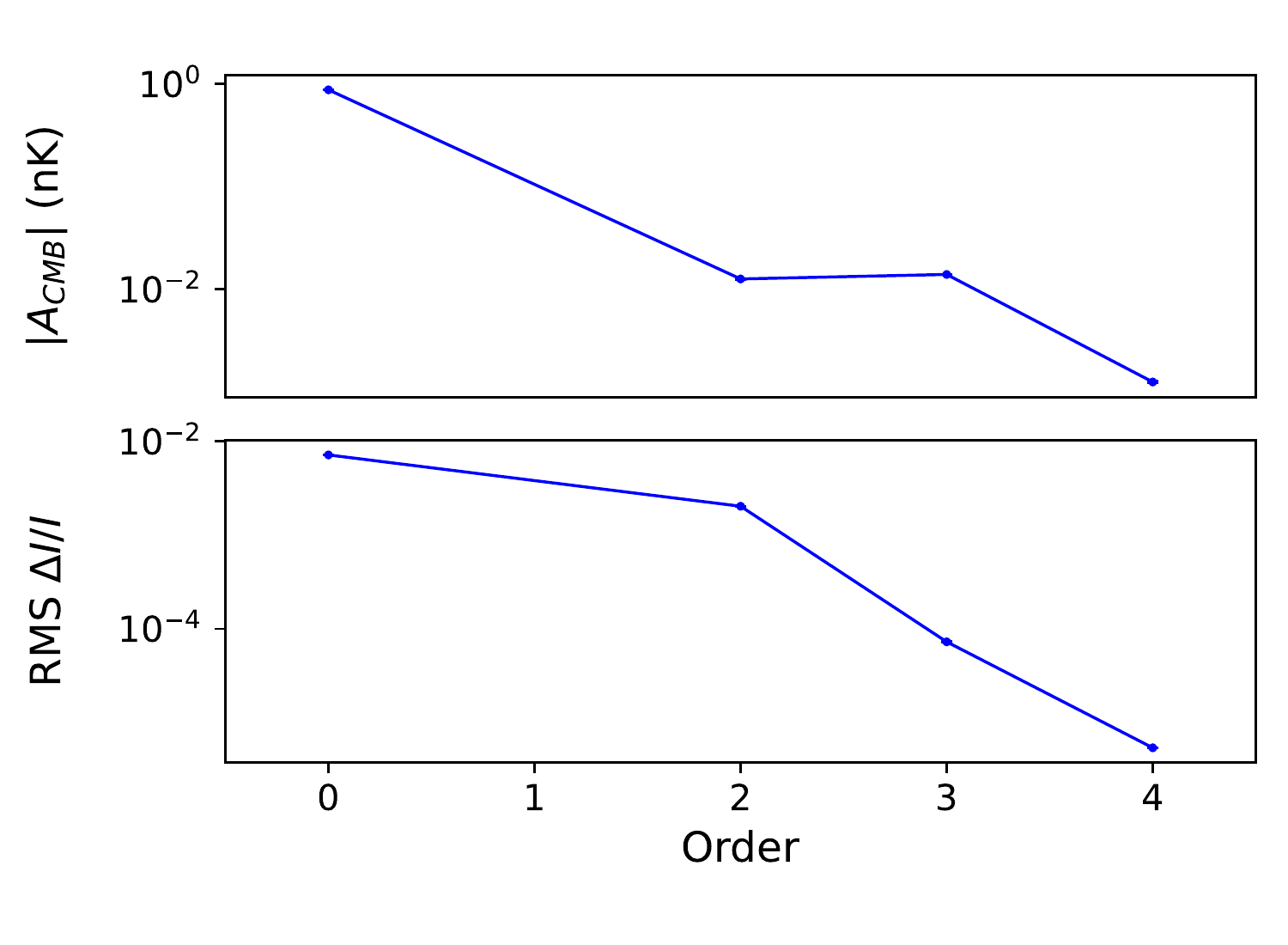}
    \includegraphics[width = 0.32\textwidth]{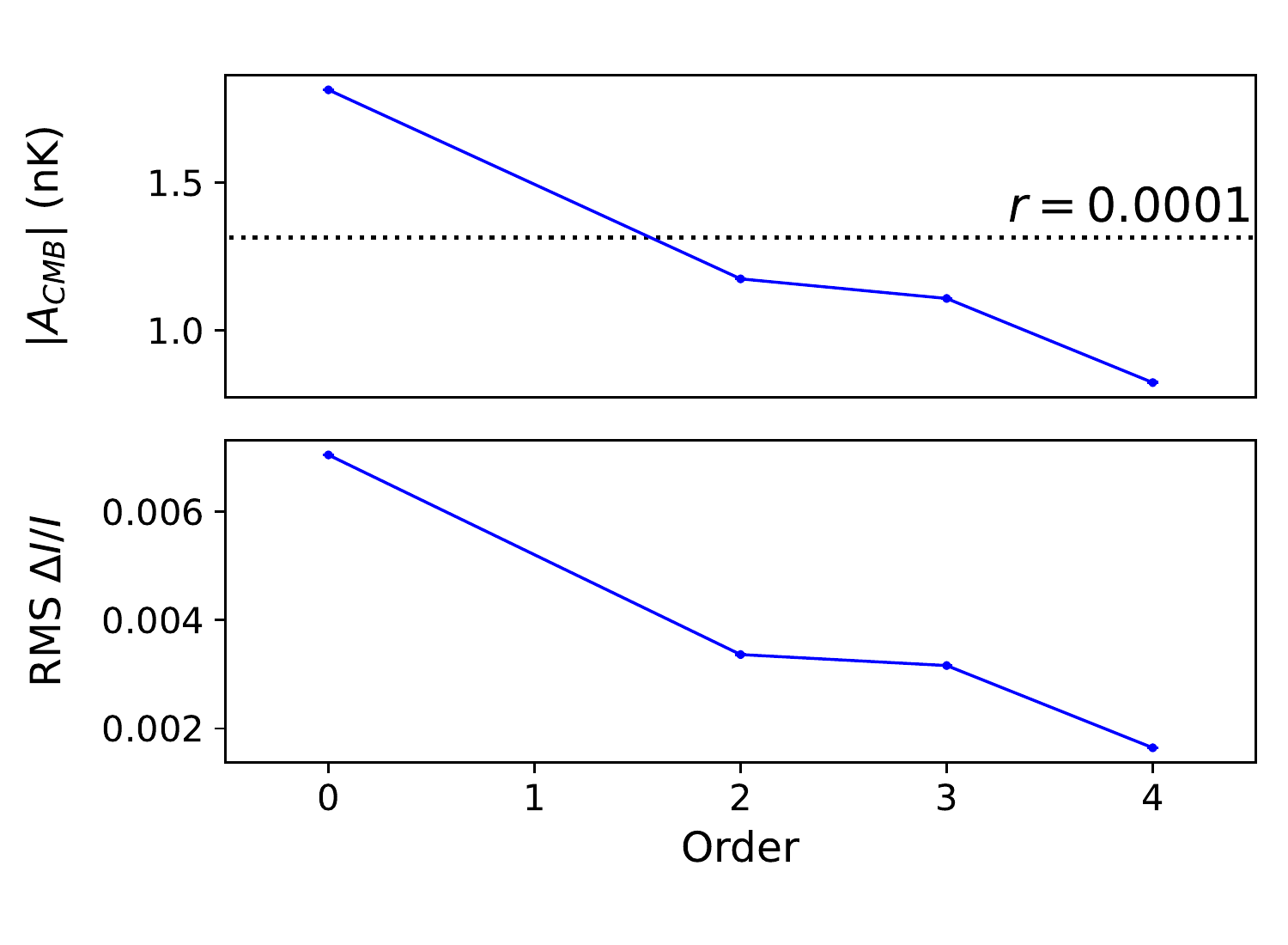}
    \caption{Best-fit CMB amplitude in nK and root-mean-square fractional residual using the One-Gaussian model (a), the Transient Heating model (b), and the Two-Gaussian model (c) for the simulated dust component.
    Horizontal lines show a corresponding $r$ value of $10^{-4}$.
    For the center panel, this line is above the upper $y$-limit.
    }
	\label{fig:bias_vs_order_PICO}
\end{figure}

Figure \ref{fig:bias_vs_order_PICO} shows the mean CMB bias and root-mean-square fractional residual as a function of expansion order using the PICO frequency bands for the One-Gaussian, Transient Heating, and Two-Gaussian models.
We choose the root-mean-square fractional residual as our goodness of fit as we are performing noiseless simulations.
Since the input models include no CMB component, the fitted CMB amplitude (top panel) represents the bias.
The root-mean-square fractional residual (bottom panel) indicates how sensitive channels need to be to detect a bias.
Note that in this case, a higher fractional residual indicates a better ability to detect bias.
We find that when using the moment method with this approach, the bias could be reduced to 
$|A_{\textrm{CMB}}| < 1$ nK ($r < 1.0 \times 10^{-4}$) in all cases without requiring prior knowledge of the true dust temperature distribution.
It is not clear why the second-order fit is worse than the zeroth-order fit for the single Gaussian model, however it is important to note that because we are expanding with respect to inverse temperature rather than temperature, two moments are not necessarily sufficient to adequately describe the distribution.

\section{Recovering the dust temperature distribution}
\label{sec:maxent}

While our primary motivation in utilizing the moment method is in minimizing bias in the tensor-to-scalar ratio by allowing for more complexity and flexibility in the dust model, this technique could potentially tell us more about the underlying distribution of dust physical parameters as well.
For second- or higher-order expansions, an approximate distribution of dust temperatures can be recovered from the truncated list of moments returned by the moment method using the principle of maximum entropy.

The classical moment problem \citep{Shohat1943} asks whether, given a sequence of moments, there is a unique probability function having those moments.
If one only has a finite number of moments, this problem is under-determined.
The idea of using a maximum entropy approach to solve an under-determined inverse problem is first introduced in \cite{Shannon1948}, and is first applied to the classical moment problem by \cite{Mead1984}.
The maximum entropy approach to the moment problem has found many applications across scientific disciplines \citep{Abramov2010}.

\begin{figure}
    \centering
    \includegraphics[width = 0.32\columnwidth]{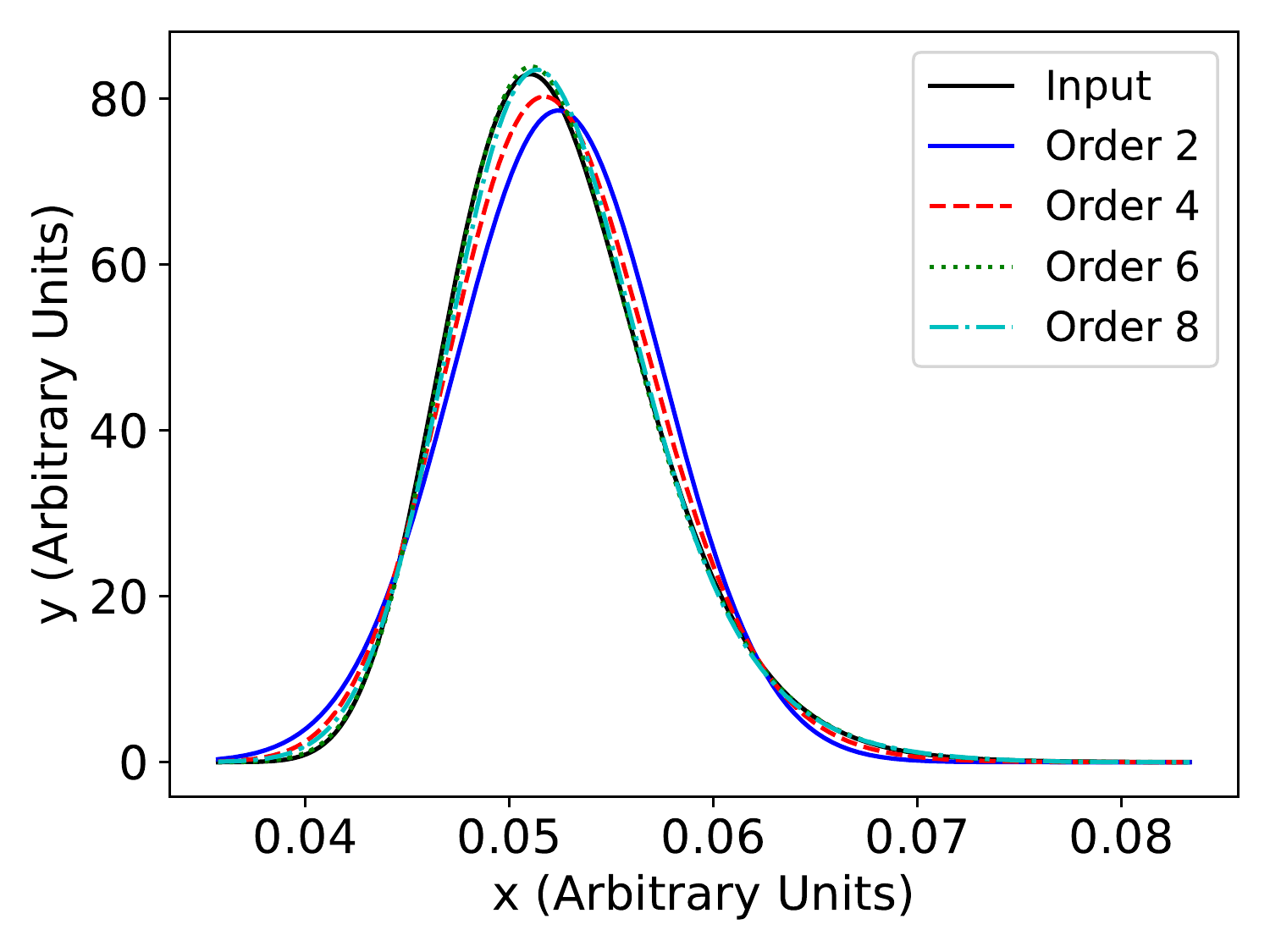}
    \includegraphics[width = 0.32\columnwidth]{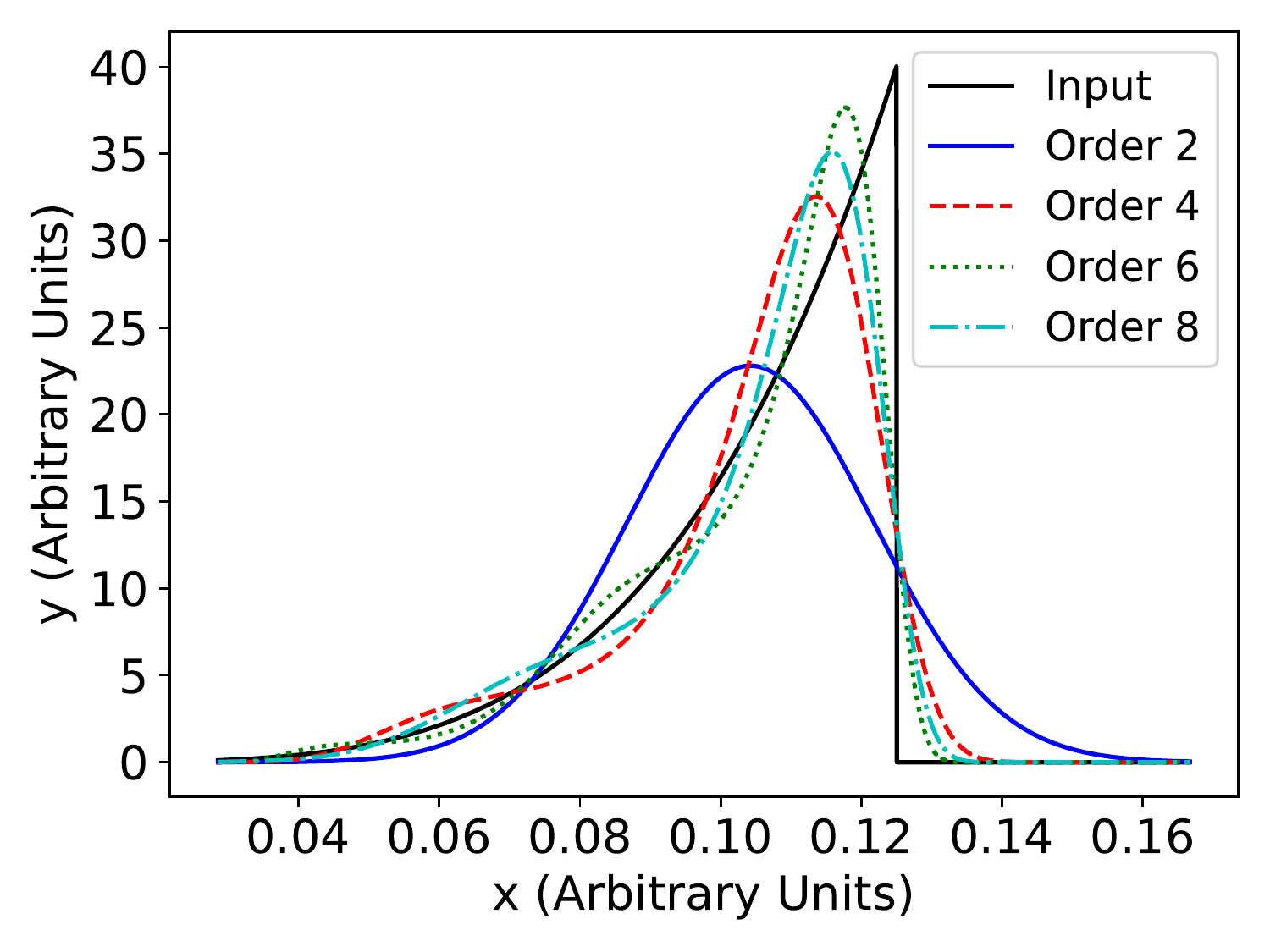}
    \includegraphics[width = 0.32\columnwidth]{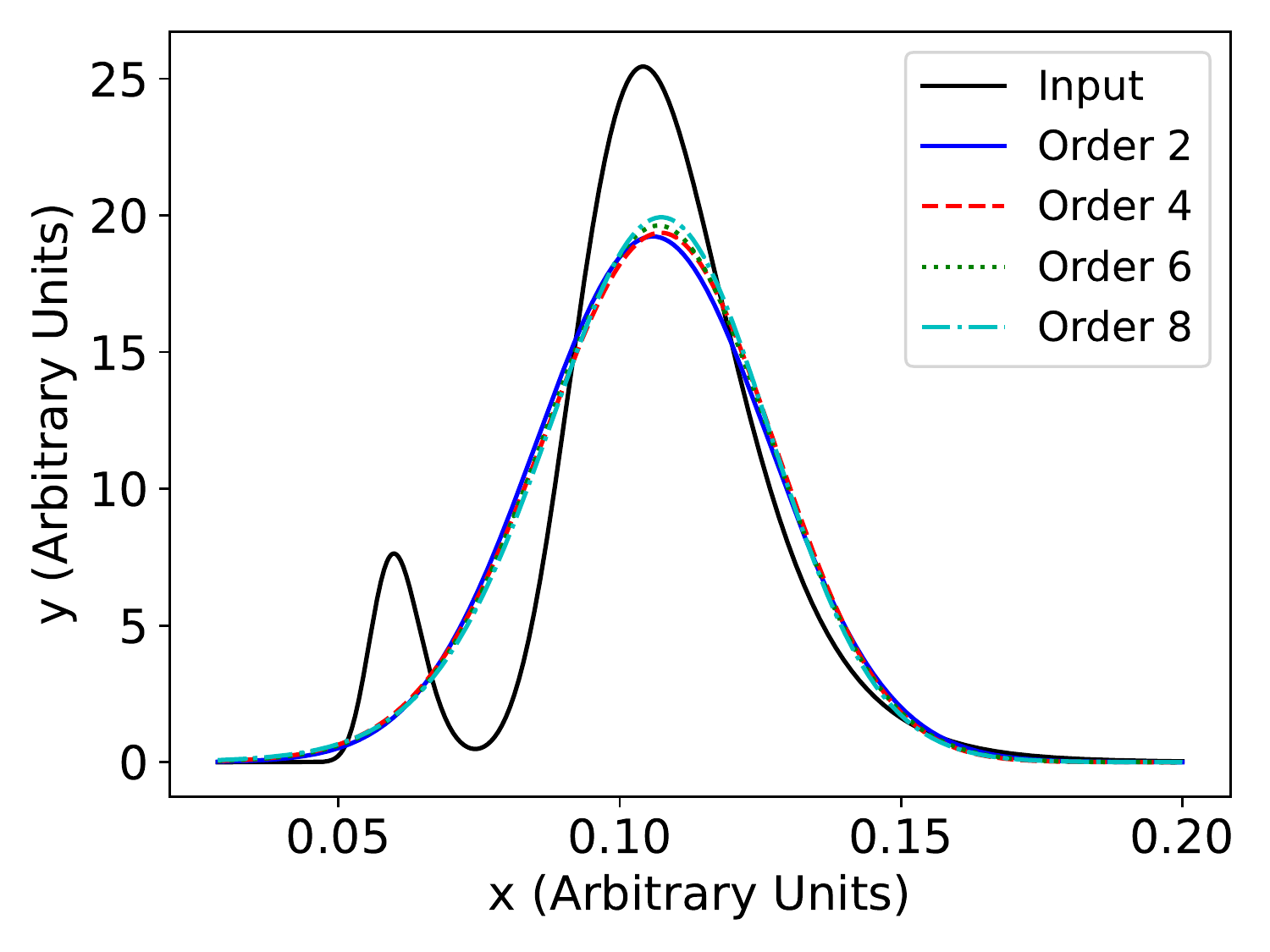}
    \caption{Recovered distributions using the maximum entropy method based on the true moments for three sample distributions based on our models.
    A distribution with two peaks would require higher-order information to adequately capture.
    }
	\label{fig:pymaxent}
\end{figure}

The principle of maximum entropy states that of all possible distributions satisfying a finite set of moment constraints, that with the largest Shannon entropy is most representative of our knowledge about the system.
\textsc{PyMaxEnt} \citep{Saad2019} is a Python package which takes a finite list of moment constraints and optimizes a parameter distribution to maximize the Shannon entropy.
This could be used to obtain an approximate distribution of dust temperatures along individual lines of sight.
Figure \ref{fig:pymaxent} shows distributions recovered from the theoretical moment values using \textsc{PyMaxEnt} as a function of the number of known moments for three sample distributions based on the considered dust models.
Note that the x-axis represents the inverse temperature, as we are expanding in terms of this parameter.

If the variation in dust properties along the line of sight is well-represented by a Gaussian distribution, two moments in the dust expansion are sufficient to recover the distribution of dust emissivity as a function of temperature along individual lines of sight.
More complex distributions require additional moments.
A single extended distribution requires expansion to fourth order of higher, while a distribution with two well-separated peaks is poorly fit at all orders.
We note, however, that the recovery process here makes no assumptions regarding the underlying dust complexity.
\cite{John2007} describe alternative approaches to recovering a distribution from a list of moments which, given additional assumptions about the distribution, can provide better recovered distributions with the available moments.

\begin{figure}
    \centering
    \includegraphics[width = 0.6\columnwidth]{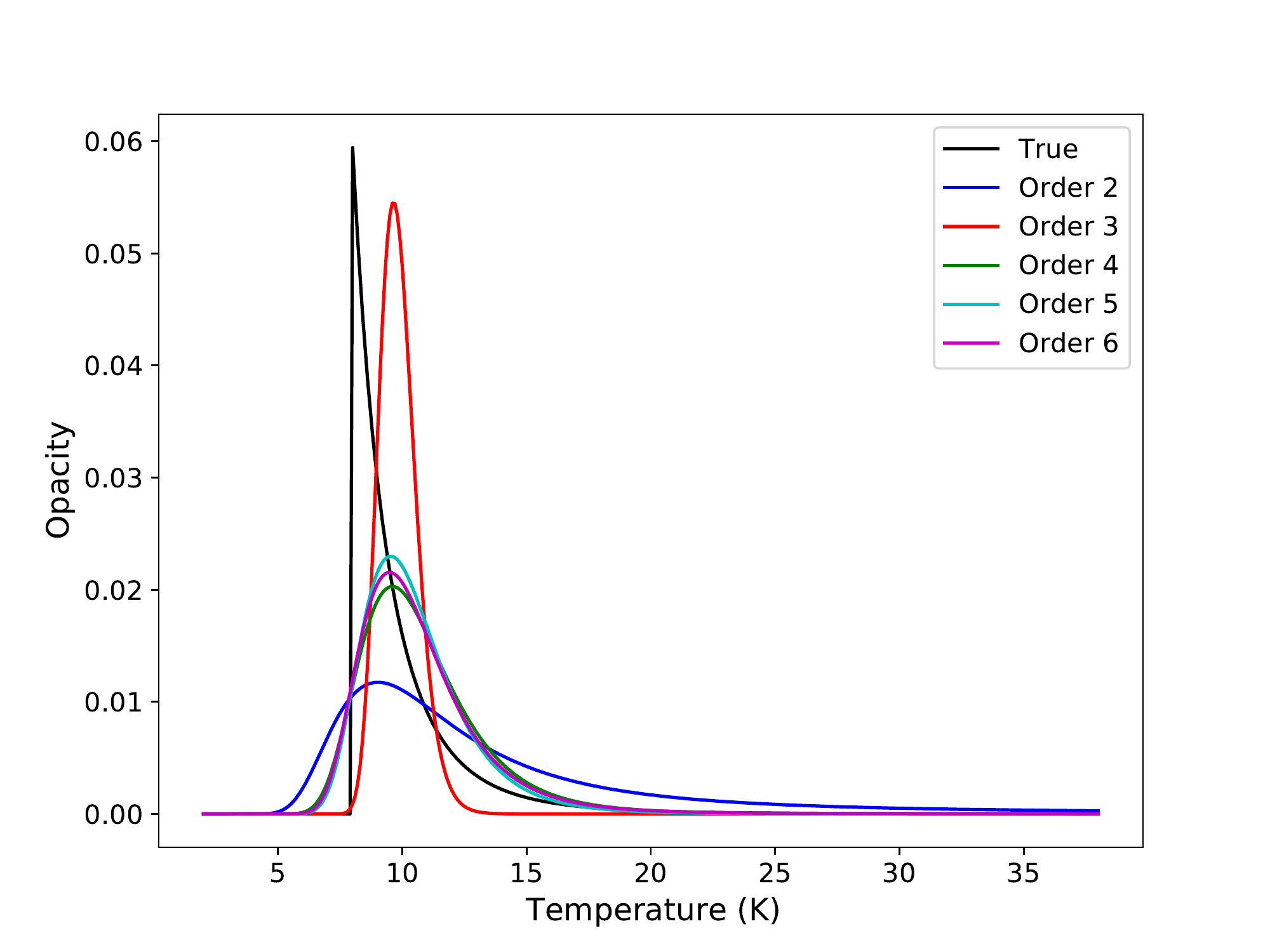}
    \caption{Recovered distribution of dust temperatures for the transient heating model with PICO}
	\label{fig:pymaxent_PICO_Transient_Heating_avg}
\end{figure}

Figure \ref{fig:pymaxent} uses the moments calculated directly from the full spectral energy distributions of the different dust models.
In practice, the frequency spectra are sampled at discrete frequencies.
Figure \ref{fig:pymaxent_PICO_Transient_Heating_avg} shows the recovered distributions from the truncated list of moments obtained by the noiseless transient heating model simulation using only the PICO frequency bands.
The lowest (second-order) expansion is sufficient to show an extended distribution in temperature, with modest improvement seen at higher order.

\section{Conclusions}
\label{sec:conclusions}

Our goal is to estimate dust complexity along the line of sight and its impact on bias in the tensor-to-scalar ratio for measurements of the B-mode polarization of the CMB.
Here, we use the scatter in dust temperature and spectral index observed for \textit{Planck} data in the plane of the sky as a proxy for similar variation along a single line of sight.
Note that every realistic beam will include at least some variations across the map as well as along the line of sight. These effects are qualitatively the same, but any realistic beam will likely have at least somewhat larger variation than a true pencil beam.
We find that an estimated line-of-sight standard deviation of 1.6K in temperature or 0.045 in spectral index will source non-trivial bias in the tensor-to-scalar ratio if fitted using a single modified blackbody model.
Bias of order $\delta r = 1.4 \times 10^{-4}$ is negligible for current upper limits to $r$ but could become significant at sensitivity levels anticipated for futuristic missions such as PICO.  
Variations in temperature are more important to capture than variations in spectral index, as the observed degree of temperature variation leads to a larger bias if not properly accounted for.

Replacing a simple modified blackbody model with an expansion in moments of the temperature distribution significantly reduces the bias without requiring \textit{a priori} knowledge of the dust physical parameters.
An expansion to second order accurately reproduces a Gaussian distribution of dust temperatures along an individual line of sight, while higher orders are required for more complicated distributions.
A futuristic experiment with information in enough frequency bands such as PICO could provide information on the dust temperature distribution along a given sightline using the statistical moments obtained from the moment method.
This could allow us to distinguish between competing dust models, shedding light on dust emission mechanisms.
That being said, as the order of the expansion is increased it is expected that there will be some trade-off between better modeling of the dust distribution and larger uncertainties in $r$.

\bibliographystyle{aasjournal}
\bibliography{bib}

\begin{thebibliography}{}
\expandafter\ifx\csname natexlab\endcsname\relax\def\natexlab#1{#1}\fi
\providecommand{\url}[1]{\href{#1}{#1}}
\providecommand{\dodoi}[1]{doi:~\href{http://doi.org/#1}{\nolinkurl{#1}}}
\providecommand{\doeprint}[1]{\href{http://ascl.net/#1}{\nolinkurl{http://ascl.net/#1}}}
\providecommand{\doarXiv}[1]{\href{https://arxiv.org/abs/#1}{\nolinkurl{https://arxiv.org/abs/#1}}}

\bibitem[{Abramov(2010)}]{Abramov2010}
Abramov, R.~V. 2010, Commun. Math. Sci., 8, 377

\bibitem[{Ade {et~al.}(2021)Ade, Ahmed, Amiri, Barkats, Thakur, Bischoff, Beck,
  Bock, Boenish, Bullock, Buza, Cheshire, Connors, Cornelison, Crumrine,
  Cukierman, Denison, Dierickx, Duband, Eiben, Fatigoni, Filippini, Fliescher,
  Goeckner-Wald, Goldfinger, Grayson, Grimes, Hall, Halal, Halpern, Hand,
  Harrison, Henderson, Hildebrandt, Hilton, Hubmayr, Hui, Irwin, Kang, Karkare,
  Karpel, Kefeli, Kernasovskiy, Kovac, Kuo, Lau, Leitch, Lennox, Megerian,
  Minutolo, Moncelsi, Nakato, Namikawa, Nguyen, O'Brient, Ogburn, Palladino,
  Prouve, Pryke, Racine, Reintsema, Richter, Schillaci, Schwarz, Schmitt,
  Sheehy, Soliman, Germaine, Steinbach, Sudiwala, Teply, Thompson, Tolan,
  Tucker, Turner, Umilt\`a, Verg\`es, Vieregg, Wandui, Weber, Wiebe, Willmert,
  Wong, Wu, Yang, Yoon, Young, Yu, Zeng, Zhang, \& Zhang}]{Ade2021}
Ade, P. A.~R., Ahmed, Z., Amiri, M., {et~al.} 2021, Phys. Rev. Lett., 127,
  151301

\bibitem[{{Andr{\'e}} {et~al.}(2014){Andr{\'e}}, {Baccigalupi}, {Banday},
  {Barbosa}, {Barreiro}, {Bartlett}, {Bartolo}, {Battistelli}, {Battye},
  {Bendo}, {Beno{\^\i}t}, {Bernard}, {Bersanelli}, {B{\'e}thermin},
  {Bielewicz}, {Bonaldi}, {Bouchet}, {Boulanger}, {Brand}, {Bucher},
  {Burigana}, {Cai}, {Camus}, {Casas}, {Casasola}, {Castex}, {Challinor},
  {Chluba}, {Chon}, {Colafrancesco}, {Comis}, {Cuttaia}, {D'Alessandro}, {Da
  Silva}, {Davis}, {de Avillez}, {de Bernardis}, {de Petris}, {de Rosa}, {de
  Zotti}, {Delabrouille}, {D{\'e}sert}, {Dickinson}, {Diego}, {Dunkley},
  {En{\ss}lin}, {Errard}, {Falgarone}, {Ferreira}, {Ferri{\`e}re}, {Finelli},
  {Fletcher}, {Fosalba}, {Fuller}, {Galli}, {Ganga}, {Garc{\'\i}a-Bellido},
  {Ghribi}, {Giard}, {Giraud-H{\'e}raud}, {Gonzalez-Nuevo}, {Grainge},
  {Gruppuso}, {Hall}, {Hamilton}, {Haverkorn}, {Hernandez-Monteagudo},
  {Herranz}, {Jackson}, {Jaffe}, {Khatri}, {Kunz}, {Lamagna}, {Lattanzi},
  {Leahy}, {Lesgourgues}, {Liguori}, {Liuzzo}, {Lopez-Caniego}, {Macias-Perez},
  {Maffei}, {Maino}, {Mangilli}, {Martinez-Gonzalez}, {Martins}, {Masi},
  {Massardi}, {Matarrese}, {Melchiorri}, {Melin}, {Mennella}, {Mignano},
  {Miville-Desch{\^e}nes}, {Monfardini}, {Murphy}, {Naselsky}, {Nati},
  {Natoli}, {Negrello}, {Noviello}, {O'Sullivan}, {Paci}, {Pagano}, {Paladino},
  {Palanque-Delabrouille}, {Paoletti}, {Peiris}, {Perrotta}, {Piacentini},
  {Piat}, {Piccirillo}, {Pisano}, {Polenta}, {Pollo}, {Ponthieu},
  {Remazeilles}, {Ricciardi}, {Roman}, {Rosset}, {Rubino-Martin}, {Salatino},
  {Schillaci}, {Shellard}, {Silk}, {Starobinsky}, {Stompor}, {Sunyaev},
  {Tartari}, {Terenzi}, {Toffolatti}, {Tomasi}, {Trappe}, {Tristram},
  {Trombetti}, {Tucci}, {Van de Weijgaert}, {Van Tent}, {Verde}, {Vielva},
  {Wandelt}, {Watson}, \& {Withington}}]{Andre2014}
{Andr{\'e}}, P., {Baccigalupi}, C., {Banday}, A., {et~al.} 2014, \jcap, 2014,
  006

\bibitem[{Armitage-Caplan {et~al.}(2012)Armitage-Caplan, Dunkley, Eriksen, \&
  Dickinson}]{ArmitageCaplan2012}
Armitage-Caplan, C., Dunkley, J., Eriksen, H.~K., \& Dickinson, C. 2012,
  Monthly Notices of the Royal Astronomical Society, 424, 1914

\bibitem[{Ashton {et~al.}(2018)Ashton, Ade, Angil{\`{e}}, Benton, Devlin,
  Dober, Fissel, Fukui, Galitzki, Gandilo, Klein, Korotkov, Li, Martin,
  Matthews, Moncelsi, Nakamura, Netterfield, Novak, Pascale, Poidevin, Santos,
  Savini, Scott, Shariff, Soler, Thomas, Tucker, Tucker, \&
  Ward-Thompson}]{Ashton2018}
Ashton, P.~C., Ade, P. A.~R., Angil{\`{e}}, F.~E., {et~al.} 2018, The
  Astrophysical Journal, 857, 10

\bibitem[{Bock {et~al.}(2008)Bock, Cooray, Hanany, Keating, Lee, Matsumura,
  Milligan, Ponthieu, Renbarger, \& Tran}]{Bock2008}
Bock, J., Cooray, A., Hanany, S., {et~al.} 2008, The Experimental Probe of
  Inflationary Cosmology (EPIC): A Mission Concept Study for NASA's Einstein
  Inflation Probe

\bibitem[{Chluba {et~al.}(2017)Chluba, Hill, \& Abitbol}]{Chluba2017}
Chluba, J., Hill, J.~C., \& Abitbol, M.~H. 2017, Monthly Notices of the Royal
  Astronomical Society, 472, 1195

\bibitem[{Draine \& Hensley(2021)}]{Draine2021}
Draine, B.~T., \& Hensley, B.~S. 2021, The Astrophysical Journal, 909, 94

\bibitem[{Eriksen {et~al.}(2004)Eriksen, O'Dwyer, Jewell, Wandelt, Larson,
  Gorski, Levin, Banday, \& Lilje}]{Eriksen2004}
Eriksen, H.~K., O'Dwyer, I.~J., Jewell, J.~B., {et~al.} 2004, The Astrophysical
  Journal Supplement Series, 155, 227

\bibitem[{Fantaye {et~al.}(2011)Fantaye, Stivoli, Grain, Leach, Tristram,
  Baccigalupi, \& Stompor}]{Fantaye2011}
Fantaye, Y., Stivoli, F., Grain, J., {et~al.} 2011, Journal of Cosmology and
  Astroparticle Physics, 2011, 001

\bibitem[{Finkbeiner {et~al.}(1999)Finkbeiner, Davis, \&
  Schlegel}]{Finkbeiner1999}
Finkbeiner, D.~P., Davis, M., \& Schlegel, D.~J. 1999, The Astrophysical
  Journal, 524, 867

\bibitem[{{Foreman-Mackey} {et~al.}(2013){Foreman-Mackey}, {Hogg}, {Lang}, \&
  {Goodman}}]{Foreman-Mackey2013}
{Foreman-Mackey}, D., {Hogg}, D.~W., {Lang}, D., \& {Goodman}, J. 2013, \pasp,
  125, 306

\bibitem[{{Guillet, V.} {et~al.}(2018){Guillet, V.}, {Fanciullo, L.},
  {Verstraete, L.}, {Boulanger, F.}, {Jones, A. P.}, {Miville-Desch\^enes,
  M.-A.}, {Ysard, N.}, {Levrier, F.}, \& {Alves, M.}}]{Guillet2018}
{Guillet, V.}, {Fanciullo, L.}, {Verstraete, L.}, {et~al.} 2018, A\&A, 610, A16

\bibitem[{{Hanany} {et~al.}(2019){Hanany}, {Alvarez}, {Artis}, {Ashton},
  {Aumont}, {Aurlien}, {Banerji}, {Barreiro}, {Bartlett}, {Basak}, {Battaglia},
  {Bock}, {Boddy}, {Bonato}, {Borrill}, {Bouchet}, {Boulanger}, {Burkhart},
  {Chluba}, {Chuss}, {Clark}, {Cooperrider}, {Crill}, {De Zotti},
  {Delabrouille}, {Di Valentino}, {Didier}, {Dor{\'e}}, {Eriksen}, {Errard},
  {Essinger-Hileman}, {Feeney}, {Filippini}, {Fissel}, {Flauger}, {Fuskeland},
  {Gluscevic}, {Gorski}, {Green}, {Hanany}, {Hensley}, {Herranz}, {Hill},
  {Hivon}, {Hlo{\v{z}}ek}, {Hubmayr}, {Johnson}, {Jones}, {Jones}, {Knox},
  {Kogut}, {L{\'o}pez-Caniego}, {Lawrence}, {Lazarian}, {Li}, {Madhavacheril},
  {Melin}, {Meyers}, {Murray}, {Negrello}, {Novak}, {O'Brient}, {Paine},
  {Pearson}, {Pogosian}, {Pryke}, {Puglisi}, {Remazeilles}, {Rocha},
  {Schmittfull}, {Scott}, {Shirron}, {Stephens}, {Sutin}, {Tomasi},
  {Trangsrud}, {van Engelen}, {Vansyngel}, {Wehus}, {Wen}, {Xu}, {Young}, \&
  {Zonca}}]{Hanany2019}
{Hanany}, S., {Alvarez}, M., {Artis}, E., {et~al.} 2019, in \baas, Vol.~51, 194

\bibitem[{Hensley \& Bull(2018)}]{Hensley2018}
Hensley, B.~S., \& Bull, P. 2018, The Astrophysical Journal, 853, 127

\bibitem[{{J. Shohat} \& {J. D. Tamarkin}(1943)}]{Shohat1943}
{J. Shohat}, \& {J. D. Tamarkin}. 1943, The Problem of Moments, Mathematical
  surveys (American Mathematical Society)

\bibitem[{John {et~al.}(2007)John, Angelov, {\"O}nc{\"u}l, \&
  Th{\'e}venin}]{John2007}
John, V., Angelov, I., {\"O}nc{\"u}l, A., \& Th{\'e}venin, D. 2007, Chemical
  Engineering Science, 62, 2890

\bibitem[{Kogut \& Fixsen(2016)}]{Kogut2016}
Kogut, A., \& Fixsen, D.~J. 2016, The Astrophysical Journal, 826, 101

\bibitem[{Kogut {et~al.}(2011)Kogut, Fixsen, Chuss, Dotson, Dwek, Halpern,
  Hinshaw, Meyer, Moseley, Seiffert, Spergel, \& Wollack}]{Kogut2011}
Kogut, A., Fixsen, D., Chuss, D., {et~al.} 2011, Journal of Cosmology and
  Astroparticle Physics, 2011, 025

\bibitem[{{Krauss} \& {Wilczek}(2014)}]{Krauss2014}
{Krauss}, L.~M., \& {Wilczek}, F. 2014, International Journal of Modern Physics
  D, 23, 1441001

\bibitem[{{Matsumura} {et~al.}(2014){Matsumura}, {Akiba}, {Borrill}, {Chinone},
  {Dobbs}, {Fuke}, {Ghribi}, {Hasegawa}, {Hattori}, {Hattori}, {Hazumi},
  {Holzapfel}, {Inoue}, {Ishidoshiro}, {Ishino}, {Ishitsuka}, {Karatsu},
  {Katayama}, {Kawano}, {Kibayashi}, {Kibe}, {Kimura}, {Kimura}, {Koga},
  {Kozu}, {Komatsu}, {Lee}, {Matsuhara}, {Mima}, {Mitsuda}, {Mizukami},
  {Morii}, {Morishima}, {Murayama}, {Nagai}, {Nagata}, {Nakamura}, {Naruse},
  {Natsume}, {Nishibori}, {Nishino}, {Noda}, {Noguchi}, {Ogawa}, {Oguri},
  {Ohta}, {Otani}, {Richards}, {Sakai}, {Sato}, {Sato}, {Sekimoto}, {Shimizu},
  {Shinozaki}, {Sugita}, {Suzuki}, {Suzuki}, {Tajima}, {Takada}, {Takakura},
  {Takei}, {Tomaru}, {Uzawa}, {Wada}, {Watanabe}, {Yoshida}, {Yamasaki},
  {Yoshida}, \& {Yotsumoto}}]{Matsumura2014}
{Matsumura}, T., {Akiba}, Y., {Borrill}, J., {et~al.} 2014, Journal of Low
  Temperature Physics, 176, 733

\bibitem[{Mead \& Papanicolaou(1984)}]{Mead1984}
Mead, L.~R., \& Papanicolaou, N. 1984, Journal of Mathematical Physics, 25,
  2404

\bibitem[{{Meny, C.} {et~al.}(2007){Meny, C.}, {Gromov, V.}, {Boudet, N.},
  {Bernard, J.-Ph.}, {Paradis, D.}, \& {Nayral, C.}}]{Meny2007}
{Meny, C.}, {Gromov, V.}, {Boudet, N.}, {et~al.} 2007, A\&A, 468, 171

\bibitem[{Osumi {et~al.}(2021)Osumi, Weiland, Addison, \& Bennett}]{Osumi2021}
Osumi, K., Weiland, J.~L., Addison, G.~E., \& Bennett, C.~L. 2021, Limits on
  Polarized Dust Spectral Index Variations for CMB Foreground Analysis

\bibitem[{{Paradis, D.} {et~al.}(2011){Paradis, D.}, {Bernard, J.-P.}, {M\'eny,
  C.}, \& {Gromov, V.}}]{Paradis2011}
{Paradis, D.}, {Bernard, J.-P.}, {M\'eny, C.}, \& {Gromov, V.} 2011, A\&A, 534,
  A118

\bibitem[{{Planck Collaboration} {et~al.}(2016){Planck Collaboration}, {Adam,
  R.}, {Ade, P. A. R.}, {Aghanim, N.}, {Arnaud, M.}, {Aumont, J.},
  {Baccigalupi, C.}, {Banday, A. J.}, {Barreiro, R. B.}, {Bartlett, J. G.},
  {Bartolo, N.}, {Battaner, E.}, {Benabed, K.}, {Benoit-L\'evy, A.}, {Bernard,
  J.-P.}, {Bersanelli, M.}, {Bielewicz, P.}, {Bonaldi, A.}, {Bonavera, L.},
  {Bond, J. R.}, {Borrill, J.}, {Bouchet, F. R.}, {Boulanger, F.}, {Bracco,
  A.}, {Bucher, M.}, {Burigana, C.}, {Butler, R. C.}, {Calabrese, E.},
  {Cardoso, J.-F.}, {Catalano, A.}, {Challinor, A.}, {Chamballu, A.}, {Chary,
  R.-R.}, {Chiang, H. C.}, {Christensen, P. R.}, {Clements, D. L.}, {Colombi,
  S.}, {Colombo, L. P. L.}, {Combet, C.}, {Couchot, F.}, {Coulais, A.}, {Crill,
  B. P.}, {Curto, A.}, {Cuttaia, F.}, {Danese, L.}, {Davies, R. D.}, {Davis, R.
  J.}, {de Bernardis, P.}, {de Zotti, G.}, {Delabrouille, J.}, {Delouis,
  J.-M.}, {D\'esert, F.-X.}, {Dickinson, C.}, {Diego, J. M.}, {Dolag, K.},
  {Dole, H.}, {Donzelli, S.}, {Dor\'e, O.}, {Douspis, M.}, {Ducout, A.},
  {Dunkley, J.}, {Dupac, X.}, {Efstathiou, G.}, {Elsner, F.}, {En\ss{}lin, T.
  A.}, {Eriksen, H. K.}, {Falgarone, E.}, {Finelli, F.}, {Forni, O.}, {Frailis,
  M.}, {Fraisse, A. A.}, {Franceschi, E.}, {Frejsel, A.}, {Galeotta, S.},
  {Galli, S.}, {Ganga, K.}, {Ghosh, T.}, {Giard, M.}, {Giraud-H\'eraud, Y.},
  {Gjerl\o{}w, E.}, {Gonz\'alez-Nuevo, J.}, {G\'orski, K. M.}, {Gratton, S.},
  {Gregorio, A.}, {Gruppuso, A.}, {Guillet, V.}, {Hansen, F. K.}, {Hanson, D.},
  {Harrison, D. L.}, {Helou, G.}, {Henrot-Versill\'e, S.},
  {Hern\'andez-Monteagudo, C.}, {Herranz, D.}, {Hivon, E.}, {Hobson, M.},
  {Holmes, W. A.}, {Huffenberger, K. M.}, {Hurier, G.}, {Jaffe, A. H.}, {Jaffe,
  T. R.}, {Jewell, J.}, {Jones, W. C.}, {Juvela, M.}, {Keih\"anen, E.},
  {Keskitalo, R.}, {Kisner, T. S.}, {Kneissl, R.}, {Knoche, J.}, {Knox, L.},
  {Krachmalnicoff, N.}, {Kunz, M.}, {Kurki-Suonio, H.}, {Lagache, G.},
  {Lamarre, J.-M.}, {Lasenby, A.}, {Lattanzi, M.}, {Lawrence, C. R.}, {Leahy,
  J. P.}, {Leonardi, R.}, {Lesgourgues, J.}, {Levrier, F.}, {Liguori, M.},
  {Lilje, P. B.}, {Linden-V\o{}rnle, M.}, {L\'opez-Caniego, M.}, {Lubin, P.
  M.}, {Mac\'{\i}as-P\'erez, J. F.}, {Maffei, B.}, {Maino, D.}, {Mandolesi,
  N.}, {Mangilli, A.}, {Maris, M.}, {Martin, P. G.}, {Mart\'{\i}nez-Gonz\'alez,
  E.}, {Masi, S.}, {Matarrese, S.}, {Mazzotta, P.}, {Meinhold, P. R.},
  {Melchiorri, A.}, {Mendes, L.}, {Mennella, A.}, {Migliaccio, M.}, {Mitra,
  S.}, {Miville-Desch\^enes, M.-A.}, {Moneti, A.}, {Montier, L.}, {Morgante,
  G.}, {Mortlock, D.}, {Moss, A.}, {Munshi, D.}, {Murphy, J. A.}, {Naselsky,
  P.}, {Nati, F.}, {Natoli, P.}, {Netterfield, C. B.}, {N\o{}rgaard-Nielsen, H.
  U.}, {Noviello, F.}, {Novikov, D.}, {Novikov, I.}, {Pagano, L.}, {Pajot, F.},
  {Paladini, R.}, {Paoletti, D.}, {Partridge, B.}, {Pasian, F.}, {Patanchon,
  G.}, {Pearson, T. J.}, {Perdereau, O.}, {Perotto, L.}, {Perrotta, F.},
  {Pettorino, V.}, {Piacentini, F.}, {Piat, M.}, {Pierpaoli, E.}, {Pietrobon,
  D.}, {Plaszczynski, S.}, {Pointecouteau, E.}, {Polenta, G.}, {Ponthieu, N.},
  {Popa, L.}, {Pratt, G. W.}, {Prunet, S.}, {Puget, J.-L.}, {Rachen, J. P.},
  {Reach, W. T.}, {Rebolo, R.}, {Remazeilles, M.}, {Renault, C.}, {Renzi, A.},
  {Ricciardi, S.}, {Ristorcelli, I.}, {Rocha, G.}, {Rosset, C.}, {Rossetti,
  M.}, {Roudier, G.}, {Rouill\'e d\'{}Orfeuil, B.}, {Rubi\~no-Mart\'{\i}n, J.
  A.}, {Rusholme, B.}, {Sandri, M.}, {Santos, D.}, {Savelainen, M.}, {Savini,
  G.}, {Scott, D.}, {Soler, J. D.}, {Spencer, L. D.}, {Stolyarov, V.},
  {Stompor, R.}, {Sudiwala, R.}, {Sunyaev, R.}, {Sutton, D.}, {Suur-Uski,
  A.-S.}, {Sygnet, J.-F.}, {Tauber, J. A.}, {Terenzi, L.}, {Toffolatti, L.},
  {Tomasi, M.}, {Tristram, M.}, {Tucci, M.}, {Tuovinen, J.}, {Valenziano, L.},
  {Valiviita, J.}, {Van Tent, B.}, {Vibert, L.}, {Vielva, P.}, {Villa, F.},
  {Wade, L. A.}, {Wandelt, B. D.}, {Watson, R.}, {Wehus, I. K.}, {White, M.},
  {White, S. D. M.}, {Yvon, D.}, {Zacchei, A.}, \& {Zonca,
  A.}}]{Planck-Collaboration2016b}
{Planck Collaboration}, {Adam, R.}, {Ade, P. A. R.}, {et~al.} 2016, A\&A, 586,
  A133

\bibitem[{{Planck Collaboration} {et~al.}(2020){Planck Collaboration}, {Akrami,
  Y.}, {Ashdown, M.}, {Aumont, J.}, {Baccigalupi, C.}, {Ballardini, M.},
  {Banday, A. J.}, {Barreiro, R. B.}, {Bartolo, N.}, {Basak, S.}, {Benabed,
  K.}, {Bernard, J.-P.}, {Bersanelli, M.}, {Bielewicz, P.}, {Bond, J. R.},
  {Borrill, J.}, {Bouchet, F. R.}, {Boulanger, F.}, {Bracco, A.}, {Bucher, M.},
  {Burigana, C.}, {Calabrese, E.}, {Cardoso, J.-F.}, {Carron, J.}, {Chiang, H.
  C.}, {Combet, C.}, {Crill, B. P.}, {de Bernardis, P.}, {de Zotti, G.},
  {Delabrouille, J.}, {Delouis, J.-M.}, {Di Valentino, E.}, {Dickinson, C.},
  {Diego, J. M.}, {Ducout, A.}, {Dupac, X.}, {Efstathiou, G.}, {Elsner, F.},
  {En\ss{}lin, T. A.}, {Falgarone, E.}, {Fantaye, Y.}, {Ferri\`ere, K.},
  {Finelli, F.}, {Forastieri, F.}, {Frailis, M.}, {Fraisse, A. A.},
  {Franceschi, E.}, {Frolov, A.}, {Galeotta, S.}, {Galli, S.}, {Ganga, K.},
  {G\'enova-Santos, R. T.}, {Ghosh, T.}, {Gonz\'alez-Nuevo, J.}, {G\'orski, K.
  M.}, {Gruppuso, A.}, {Gudmundsson, J. E.}, {Guillet, V.}, {Handley, W.},
  {Hansen, F. K.}, {Herranz, D.}, {Huang, Z.}, {Jaffe, A. H.}, {Jones, W. C.},
  {Keih\"anen, E.}, {Keskitalo, R.}, {Kiiveri, K.}, {Kim, J.}, {Krachmalnicoff,
  N.}, {Kunz, M.}, {Kurki-Suonio, H.}, {Lamarre, J.-M.}, {Lasenby, A.}, {Le
  Jeune, M.}, {Levrier, F.}, {Liguori, M.}, {Lilje, P. B.}, {Lindholm, V.},
  {L\'opez-Caniego, M.}, {Lubin, P. M.}, {Ma, Y.-Z.}, {Mac\'{\i}as-P\'erez, J.
  F.}, {Maggio, G.}, {Maino, D.}, {Mandolesi, N.}, {Mangilli, A.}, {Martin, P.
  G.}, {Mart\'{\i}nez-Gonz\'alez, E.}, {Matarrese, S.}, {McEwen, J. D.},
  {Meinhold, P. R.}, {Melchiorri, A.}, {Migliaccio, M.}, {Miville-Desch\^enes,
  M.-A.}, {Molinari, D.}, {Moneti, A.}, {Montier, L.}, {Morgante, G.}, {Natoli,
  P.}, {Pagano, L.}, {Paoletti, D.}, {Pettorino, V.}, {Piacentini, F.},
  {Polenta, G.}, {Puget, J.-L.}, {Rachen, J. P.}, {Reinecke, M.}, {Remazeilles,
  M.}, {Renzi, A.}, {Rocha, G.}, {Rosset, C.}, {Roudier, G.},
  {Rubi\~no-Mart\'{\i}n, J. A.}, {Ruiz-Granados, B.}, {Salvati, L.}, {Sandri,
  M.}, {Savelainen, M.}, {Scott, D.}, {Soler, J. D.}, {Spencer, L. D.},
  {Tauber, J. A.}, {Tavagnacco, D.}, {Toffolatti, L.}, {Tomasi, M.},
  {Trombetti, T.}, {Valiviita, J.}, {Vansyngel, F.}, {Van Tent, B.}, {Vielva,
  P.}, {Villa, F.}, {Vittorio, N.}, {Wehus, I. K.}, {Zacchei, A.}, \& {Zonca,
  A.}}]{Planck-Collaboration2020a}
{Planck Collaboration}, {Akrami, Y.}, {Ashdown, M.}, {et~al.} 2020, A\&A, 641,
  A11

\bibitem[{{Pollack} {et~al.}(1994){Pollack}, {Hollenbach}, {Beckwith},
  {Simonelli}, {Roush}, \& {Fong}}]{Pollack1994}
{Pollack}, J.~B., {Hollenbach}, D., {Beckwith}, S., {et~al.} 1994, \apj, 421,
  615

\bibitem[{Remazeilles {et~al.}(2016)Remazeilles, Dickinson, Eriksen, \&
  Wehus}]{Remazeilles2016}
Remazeilles, M., Dickinson, C., Eriksen, H. K.~K., \& Wehus, I.~K. 2016,
  Monthly Notices of the Royal Astronomical Society, 458, 2032

\bibitem[{Remazeilles {et~al.}(2020)Remazeilles, Rotti, \&
  Chluba}]{Remazeilles2020}
Remazeilles, M., Rotti, A., \& Chluba, J. 2020, Peeling off foregrounds with
  the constrained moment ILC method to unveil primordial CMB $B$-modes

\bibitem[{Rotti \& Chluba(2020)}]{Rotti2020}
Rotti, A., \& Chluba, J. 2020, Monthly Notices of the Royal Astronomical
  Society, 500, 976

\bibitem[{Saad \& Ruai(2019)}]{Saad2019}
Saad, T., \& Ruai, G. 2019, SoftwareX, 10, 100353

\bibitem[{Shannon(1948)}]{Shannon1948}
Shannon, C.~E. 1948, The Bell System Technical Journal, 27, 379

\bibitem[{Stompor {et~al.}(2016)Stompor, Errard, \& Poletti}]{Stompor2016}
Stompor, R., Errard, J., \& Poletti, D. 2016, Phys. Rev. D, 94, 083526

\bibitem[{{The COrE Collaboration} {et~al.}(2011){The COrE Collaboration},
  Armitage-Caplan, Avillez, Barbosa, Banday, Bartolo, Battye, Bernard,
  de~Bernardis, Basak, Bersanelli, Bielewicz, Bonaldi, Bucher, Bouchet,
  Boulanger, Burigana, Camus, Challinor, Chongchitnan, Clements, Colafrancesco,
  Delabrouille, De~Petris, De~Zotti, Dickinson, Dunkley, Ensslin, Fergusson,
  Ferreira, Ferriere, Finelli, Galli, Garcia-Bellido, Gauthier, Haverkorn,
  Hindmarsh, Jaffe, Kunz, Lesgourgues, Liddle, Liguori, Lopez-Caniego, Maffei,
  Marchegiani, Martinez-Gonzalez, Masi, Mauskopf, Matarrese, Melchiorri,
  Mukherjee, Nati, Natoli, Negrello, Pagano, Paoletti, Peacocke, Peiris,
  Perroto, Piacentini, Piat, Piccirillo, Pisano, Ponthieu, Rath, Ricciardi,
  Martin, Salatino, Shellard, Stompor, Urrestilla, Van~Tent, Verde, Wandelt, \&
  Withington}]{COrE-Collaboration2011}
{The COrE Collaboration}, Armitage-Caplan, C., Avillez, M., {et~al.} 2011, COrE
  (Cosmic Origins Explorer) A White Paper

\bibitem[{{Vacher, L.} {et~al.}(2022){Vacher, L.}, {Aumont, J.}, {Montier, L.},
  {Azzoni, S.}, {Boulanger, F.}, \& {Remazeilles, M.}}]{Vacher2022}
{Vacher, L.}, {Aumont, J.}, {Montier, L.}, {et~al.} 2022, A\&A, 660, A111

\end{thebibliography}

\end{document}